\documentclass[12pt]{article}
\usepackage{amsmath,amssymb,mathrsfs,url,graphicx}
\usepackage{cite}
\usepackage[pdftex]{hyperref}
\usepackage{color}

\title{Chaotic Bohmian trajectories for stationary states}

\author{Alexandre Cesa\footnote{Institut de Physique Nucl\'eaire, Atomique et de Spectroscopie, CESAM, Universit\'e de Li\`ege, B\^at.\ B15, B - 4000 Li\`ege, Belgium. E-mail: a.cesa@ulg.ac.be}, John Martin\footnote{Institut de Physique Nucl\'eaire, Atomique et de Spectroscopie, CESAM, Universit\'e de Li\`ege, B\^at.\ B15, B - 4000 Li\`ege, Belgium. E-mail: jmartin@ulg.ac.be}, and Ward Struyve\footnote{Mathematisches Institut, Ludwig-Maximilians-Universit\"at M\"unchen, Theresienstr.\ 39, 80333 M\"unchen, Germany. E-mail: ward.struyve@gmail.com}
}

\def\pa{\partial}

\def\ii{\textrm i}
\def\ee{\textrm e}

\newcommand{\be}{\begin{equation}}
\newcommand{\en}{\end{equation}}

\begin{document}
\maketitle

\begin{abstract}
\noindent
In Bohmian mechanics, the nodes of the wave function play an important role in the generation of chaos. However, so far, most of the attention has been on moving nodes; little is known about the possibility of chaos in the case of stationary nodes. We address this question by considering stationary states, which provide the simplest examples of wave functions with stationary nodes. We provide examples of stationary wave functions for which there is chaos, as demonstrated by numerical computations, for one particle moving in 3 spatial dimensions and for two and three entangled particles in two dimensions. Our conclusion is that the motion of the nodes is not necessary for the generation of chaos. What is important is the overall complexity of the wave function. That is, if the wave function, or rather its phase, has complex spatial variations, it will lead to complex Bohmian trajectories and hence to chaos. Another aspect of our work concerns the average Lyapunov exponent, which quantifies the overall amount of chaos. Since it is very hard to evaluate the average Lyapunov exponent analytically, which is often computed numerically, it is useful to have simple quantities that agree well with the average Lyapunov exponent. We investigate possible correlations with quantities such as the participation ratio and different measures of entanglement, for different systems and different families of stationary wave functions. We find that these quantities often tend to correlate to the amount of chaos. However, the correlation is not perfect, because, in particular, these measures do not depend on the form of the basis states used to expand the  wave function,  while the amount of chaos does.
\end{abstract}

\section{Introduction}
A dynamical system is chaotic if it displays high sensitivity to initial conditions. In classical mechanics, this manifests itself by an exponential divergence of initially nearby trajectories in phase space. In quantum mechanics, the unitary evolution of the wave function preserves distances in Hilbert space and therefore prevents high sensitivity to initial conditions, so that there is no chaos in this sense. Instead, quantum chaos refers to other generic features of quantum systems,  such as energy spectra exhibiting level repulsion or ergodic eigenstates~\cite{haak01}. For example, quantum systems whose classical limit is chaotic display those signatures of quantum chaos. This paper does not concern quantum chaos, but rather chaos appearing in the framework of Bohmian mechanics. Bohmian mechanics is an alternative to standard quantum mechanics which describes actual point-particles moving under the influence of the wave function \cite{bohm93,holland93b,duerr09,duerr12}. Since there are actual trajectories, one can investigate their chaotic properties in the same way as in classical mechanics, bearing in mind that for Bohmian mechanics the relevant dynamical space is configuration space instead of phase space \cite{duerr92c,schwengelbeck95}. There is a large body of literature on chaos in Bohmian mechanics, see e.g.\ \cite{cushing00,efthymiopoulos06} for an overview. In particular, it has been shown that chaos can be obtained for a Bohmian system even though the corresponding classical system exhibits no chaos \cite{parmenter95,depolavieja96c,dealcantara98,efthymiopoulos06}. This is not so surprising, considering the fact that in Bohmian mechanics the dynamics is entirely determined by the wave function. Even though the (quantum) Hamiltonian determines the evolution of the wave function, the wave function generically leads to a chaotic dynamics for the point-particles. It has also been shown that Bohmian chaos does not necessarily correspond to quantum chaos \cite{contopoulos08,matzkin08}. The study of chaos provides further insight into the nature of the Bohmian trajectories, which is particularly important in view of the fact that the details of the Bohmian trajectories can be revealed by experiments \cite{kocsis11,mahler16}. Establishing chaos usually also implies ergodicity, which means uniqueness of the quantum equilibrium distribution. This adds further evidence for its use as a measure of typicality \cite{duerr92a,goldstein07}. 

Previous studies have shown that nodes of the wave functions (i.e., points where the wave function is zero) play an important role in the generation of chaos \cite{frisk97,konkel98,wu99,falsaperla03,wisniacki05,wisniacki06,efthymiopoulos07,contopoulos08,efthymiopoulos09,borondo09,contopoulos12}. In particular, the amount of chaos usually tends to increase with the number of nodes \cite{frisk97,wisniacki06}. Perhaps the most detailed study is in \cite{efthymiopoulos07,contopoulos08,efthymiopoulos09,contopoulos12}. There it was shown that for 2-d systems, the presence of a so-called $X$-point near the node gives rise to chaos \cite{efthymiopoulos07,contopoulos08,efthymiopoulos09}. Most of these works concern the study of moving nodes. Little is known about the presence of chaos in the case of stationary nodes. 

The aim of the paper is to further investigate the generation of chaos in the case of stationary nodes. The simplest examples of wave functions with stationary nodes are stationary states, i.e., wave functions of the form $\psi(x,t) = \ee^{-\ii Et}\,\phi(x)$ (since the density $|\psi|^2$ is stationary the nodes must be stationary as well).\footnote{It might be possible to have a non-stationary state which admits stationary nodes. It would be interesting to see if such states allow for chaotic motion. We have found no example of such a wave function. For this question, we considered the Coulomb potential for which the excited energy eigenstates all have a node at the origin. However, we were unable to find a superposition of such states that has no other nodes.} We show with various examples (concerning different Hamiltonians and wave functions) that motion of the nodes is not necessary to generate chaos (contrary to what seems to be claimed in \cite{wisniacki05}). More precisely, we give examples of stationary wave functions for which numerical simulations show that the Bohmian dynamics is chaotic (as quantified by the maximal Lyapunov exponent) for one particle moving in a harmonic potential in 3 spatial dimensions in section \ref{1particle}, and for two and three entangled particles in a 2-d square box and in a 2-d harmonic potential respectively in sections \ref{2particle} and \ref{3particle}. This shows that the nodes do not need to move in order to generate chaos. What is important is the overall complexity of the wave function, or rather its phase. Since the Bohmian velocity is proportional to the gradient of the phase, a phase which has complex variations over space will determine a complex velocity field and hence a complex dynamics, possibly leading to chaotic motion. 

Stationary states were considered before in a number of papers \cite{frisk97,parmenter97,schlegel08}. In \cite{parmenter97,schlegel08}, examples of 4-d systems were considered (corresponding to two particles in 2-d) for which there is chaos. It was argued that chaos was not possible for a stationary state in 3-d~\cite{parmenter97}. However, as we will explain in section \ref{generalconsiderations}, the argument in~\cite{parmenter97} is not correct. This will also be illustrated by the examples in section \ref{1particle}. Since a stationary state leads to an autonomous Bohmian dynamics, the Poincar\'e-Bendixson theorem\footnote{The Poincar\'e-Bendixson theorem states that if a trajectory is confined to a closed, bounded region of the plane, with no fixed points, then it must approach a closed orbit, hence it cannot be chaotic.} implies that the lowest possible dimension for which chaos is possible for a stationary state is three (cf.\ section \ref{generalconsiderations}). In \cite{frisk97}, it was claimed that a particular stationary state for a particle in 3-d yields chaos in certain regions of configuration space. However, as we explain in section \ref{generalconsiderations}, this claim is unfounded as in this case, there are constants of motion which make the motion regular. 

On the other hand, moving nodes are of course no guarantee for chaos either. In section \ref{nochaos}, we give some simple examples. 

Another aspect of our work concerns the average Lyapunov exponent, which quantifies the overall amount of chaos. Since it is very hard to evalute the average Lyapunov exponent analytically, which is often computed numerically, it is useful to have simple quantities that agree well with the average Lyapunov exponent. In sections \ref{1particle} and \ref{3particle}, we compare the average Lyapunov exponent with the participation ratio. The participation ratio (PR), recall in section \ref{measures}, measures how much a state is ``delocalized'' in a given basis, i.e., from how much basis states the state is made up. Clearly, generically, the more terms in a superposition, the more complex the Bohmian motion and hence the more possibility of chaos. The PR is a widely used measure of delocalization of wave functions in the context of quantum chaos \cite{brown08}. A quantum chaotic system will typically  be characterized by an ergodic wave function with large PR, increasing linearly with the system size (dimension of the Hilbert space).
In section~\ref{3particle}, we also compare the average Lyapunov exponent to different measures of entanglement. We consider the Meyer-Wallach measure of entanglement, the geometric measure of entanglement and the three-tangle, which we define in section \ref{measures}. For separable (i.e., non-entangled) states, the dynamics of the different particles is decoupled and thus the possibility of chaos is the same as in the single-particle case. Hence, entanglement usually increases the complexity of the Bohmian dynamics. 

We find that these measures often tend to relate to the amount of chaos. However, there are shortcomings of these measures. For starters, these measures do not depend on the form of the basis states used to expand the wave function, while the amount of chaos does. In addition, the participation ratio does not depend on the relative phases that appear in the coefficients of the expansion terms in that basis.

In appendix \ref{chaos}, we give details about how we quantify chaos as well as how we computed the Lyapunov exponent. In appendix \ref{energyeigenstates}, we recall the energy eigenstates for the considered systems.

\section{General considerations}\label{generalconsiderations}
For simplicity, we put the masses and charges of the particles as well as $\hbar$ equal to one.

\subsection{Bohmian mechanics}
Bohmian mechanics describes actual point-particles moving under the influence of the wave function \cite{bohm93,holland93b,duerr09,duerr12}. Denoting the particle positions by ${\bf x}_k$, $k=1,\dots,n$, and the configuration by $x = ({\bf x}_1, \dots, {\bf x}_n)$, the dynamics is given by the guidance equation
\be
\frac{d x}{ dt} = v^\psi(x,t) ,
\label{1}
\en
where the velocity field $v^\psi = ({\bf v}^\psi_1,\dots,{\bf v}^\psi_n)$ is given by
\be
{\bf v}_k^\psi(x,t) = \frac{\hbar}{m_k} \Im\left(\frac{\boldsymbol{\nabla}_k\psi(x,t)}{\psi( x,t)}\right) = \frac{\hbar}{m_k} \boldsymbol{\nabla}_kS (x,t),
\label{2}
\en
with $\psi = |\psi|\ee^{\ii S}$ and $m_k$ the mass of the $k$-th particle. The wave function $\psi$ satisfies the usual Schr\"odinger equation. This theory reproduces the predictions of standard quantum theory provided that for an ensemble of systems all with the same wave function, the particle configuration is distributed according to $|\psi(x,0)|^2$. This will  be the case for typical initial configurations of the universe \cite{duerr09,duerr12}.

\subsection{Stationary states}
If the spectrum is non-degenerate then the energy eigenstates can be chosen real and the Bohmian particles do not move (since then $S=0$). So in order to allow for chaos we need to consider a degenerate spectrum. We will consider stationary superpositions of the form 
\be
\psi({\bf x},t) = \ee^{-\ii Et}  \phi({\bf x}) =  \ee^{-\ii Et} \sum_i \phi_i({\bf x}),
\label{3}
\en
where the energy eigenstates $\phi_i$ are eigenstates of a complete set of commuting observables so that they constitute an orthonormal basis of the Hilbert space. 

We will consider two types of states $\phi_i$. The first type is of the form 
\be
\phi({\bf x}) = f(x)g(y)h(z),
\label{5}
\en
with ${\bf x}=(x,y,z)$ and $f$, $g$ and $h$ real functions. Such states will be considered for the 3d harmonic oscillator ($\phi^{\textrm{3d}}_{n_x,n_y,n_z}$). We will also consider systems in 2-d such as the particle in a 2d-box ($\phi^{\textrm{box}}_{n_x,n_y}$) and the 2d harmonic oscillator ($\phi^{\textrm{2d}}_{n_x,n_y}$). In that case the $z$-coordinate should be dropped. 
The second type is of the form
\be
\phi({\bf x}) = f(r)g(\theta) \ee^{\ii m \varphi},
\label{6}
\en
where $(r,\theta,\varphi)$ are spherical coordinates, with $\theta$ the polar angle and $\varphi$ the azimuthal angle, and $f$ and $g$ real. These states are eigenstates of the angular momentum operator ${\widehat L}_z = -\ii \pa_\phi$ with eigenvalue $m$. Such states will be considered for the 3d harmonic oscillator ($\phi^{\textrm{sph}}_{k,l,m}$). The explicit expressions of the states are given in appendix B. In terms of spherical coordinates the guidance equation reads
\be
\frac{dr}{dt} = \frac{\pa S}{\pa r} \,, \qquad \frac{d\theta}{dt} = \frac{1}{r^2} \frac{\pa S}{\pa \theta} \,, \qquad  \frac{d\varphi}{dt} = \frac{1}{r^2 \sin^2\theta}\frac{\pa S}{\pa \varphi}.
\label{7}
\en
For systems in two dimensions and using polar coordinates $(r,\varphi)$, one should put $\theta=\pi/2$, like for the 2d harmonic oscillator ($\phi^{\textrm{pol}}_{n_r,n_l}$).

\subsection{Constants of motion}\label{cst_of_motion}
For chaotic motion to be possible, the number of dimensions needs to be at least 3. This follows from the Poincar\'e-Bendixson theorem~\cite{lichtenberg92}. For a non-stationary dynamics this number needs to be at least 2 (since such a dynamics is equivalent to a stationary one by introducing an independent variable $\tau$, treating time as a dependent variable $t(\tau)$ and introducing the equation of motion $dt/d\tau = 1$). For a stationary wave function the velocity field is stationary too. Hence for such a state we need at least 3 spatial dimensions for chaos to be possible.

Constants of motion reduce the effective number of degrees of freedom. For example, for states of the form (\ref{6}), the phase does not depend on $r$ and $\theta$ and hence the guidance equation \eqref{7} implies that $r$ and $\theta$ are constant. The only effective degree of freedom is $\varphi$. The possible trajectories are circles \cite{holland93b} and there is no chaos. 

There are a couple of types of constant of motion that we will encounter. The first type arises for a superposition of the form
\be
\psi(x_1,\dots,x_n,t) = f_1(x_1) g_1(x_2) \chi_1(x_3,\dots,x_n,t) +  f_2(x_1) g_2(x_2) \chi_2(x_3,\dots,x_n,t),
\label{8}
\en
with $x_1,\dots,x_n$ Cartesian coordinates, and where $f_i$ and $g_i$, $i=1,2$, are real. For such a wave function, the guidance equations for $x_1$ and $x_2$ are of the form 
\be
\frac{dx_1}{dt} = (f_1 \pa_{x_1} f_2 - f_2 \pa_{x_1} f_1 )g_1 g_2 \,  \frac{\Im (\chi^*_1 \chi_2)}{|\psi|^2} , \quad \frac{dx_2}{dt} = (g_1 \pa_{x_2} g_2 - g_2 \pa_{x_2} g_1 ) f_1 f_2 \, \frac{\Im (\chi^*_1 \chi_2)}{|\psi|^2} .
\en
Dividing these velocity components and separating the variables $x_1$ and $x_2$, we obtain
\be
f(x_1)\frac{dx_1}{dt} = g(x_2)\frac{dx_2}{dt},
\en
with
\begin{equation}
f=\frac{f_1f_2}{f_1\partial_{x_1} f_2 - f_2\partial_{x_1} f_1 } , \qquad g=\frac{g_1g_2}{g_1\partial_{x_2} g_2 - g_2\partial_{x_2} g_1 }.
\end{equation}
Integration over time yields the constant of motion
\begin{equation}
C(x_1,x_2)= F({x}_1)-G({x}_2) ,
\label{com}
\end{equation}
with $dF(x_1)/dx_1=f(x_1)$ and $dG(x_2)/dx_2=g(x_2)$.

In \cite{frisk97}, two energy eigenstates of the form $\phi^{\textrm{box}}_{n_x,n_y,n_z} + \ii \phi^{\textrm{box}}_{n_y,n_z,n_x}$ where considered for a particle in a square box. It was claimed that for $(n_x,n_y,n_z) = (3,2,1)$, the dynamics was regular, while for $(n_x,n_y,n_z) = (103,102,101)$, the dynamics was chaotic in certain regions. However, there are two independent constants of motion of the form \eqref{com} which depend on different pairs of coordinates, so that there is only one effective degree of freedom and there cannot be chaos, regardless of the choice of quantum numbers $(n_x,n_y,n_z)$.

For spherical coordinates, we can have similar constants of motion (examples are found in \cite{colijn04}). For example, consider a superposition of the form
\be
\psi(r_1,\theta_1,\varphi_1,\dots,r_n,\theta_n,\varphi_n,t) = f_1(r_1) g_1(\theta_1) \chi_1(\varphi_1,\dots,t) +  f_2(r_1) g_2(\theta_1) \chi_2(\varphi_1,\dots,t),
\en
where $f_i$ and $g_i$ are real, $i=1,2$. The guidance equations for $r_1$ and $\theta_1$ are of the form 
\be
\frac{dr_1}{dt} = (f_1 \pa_{r_1} f_2 - f_2 \pa_{r_1} f_1 ) g_1 g_2\, \frac{\Im (\chi^*_1 \chi_2)}{|\psi|^2} , \quad \frac{d\theta_1}{dt} =  \frac{1}{r^2_1} (g_1 \pa_{\theta_1} g_2 - g_2 \pa_{\theta_1} g_1 )f_1 f_2\,  \frac{\Im (\chi^*_1 \chi_2)}{|\psi|^2} .
\en
In this case, there is the constant of motion 
\begin{equation}
C(r_1,\theta_1)= F({r}_1)-G({\theta}_1),
\label{com2}
\end{equation}
with $dF(r_1)/dr_1=f(r_1)$ and $dG(\theta_1)/d\theta_1=g(\theta_1)$, and
\begin{equation}
f=\frac{1}{r^2_1} \frac{f_1f_2}{f_1\partial_{r_1} f_2 - f_2\partial_{r_1} f_1 } , \qquad g=\frac{g_1g_2}{g_1\partial_{\theta_1} g_2 - g_2\partial_{\theta_1} g_1 }.
\end{equation}

Finally, consider a superposition of the form
\begin{multline}
\psi(x_1,\dots,x_n,t) = f_1(x_1) f_2(x_2) f_2(x_3)\chi_1(x_4,\dots,x_n,t)\\ +  f_2(x_1) f_1(x_2) f_2(x_3) \chi_2(x_4,\dots,x_n,t)+f_2(x_1) f_2(x_2) f_1(x_3) \chi_3(x_4,\dots,x_n,t),
\label{163}
\end{multline}
with $x_1,\dots,x_n$ Cartesian coordinates, and where $f_1$ and $f_2$, are real. One can show similarly as above that there is the constant of motion
\begin{equation}
C(x_1,x_2,x_3)= F({x}_1)+F({x}_2)+F({x}_3),
\label{com3}
\end{equation}
with $dF(x)/dx=f(x)$ and
\begin{equation}
f=\frac{f_1f_2}{f_1\partial_{x} f_2 - f_2\partial_{x} f_1}.
\end{equation}

It is often reported that for a stationary state, the energy of the Bohmian particles
\be
{\mathcal E}(x,t) = \sum_k \frac{|{\bf v}^\psi_k (x,t)|^2 }{2} + V(x) + Q(x,t)
\label{15}
\en
is a constant of motion, see e.g.\ \cite{parmenter97,schlegel08}. The quantity $V$ is the usual potential energy and $Q$ is the {\em quantum potential}, defined as $Q = - \sum_k \nabla^2_k |\psi| / 2 |\psi|$. It appears as an extra potential in Newton's equation 
\be
\frac{d^2 {\bf x}_k}{ dt^2} = - {\boldsymbol \nabla}_k (V +Q) ,
\label{16}
\en
which follows from taking the time derivative of the guidance equation. Along a trajectory, we have that $d {\mathcal E} / dt = \pa Q / \pa t$, so that ${\mathcal E}$ is generically not conserved. For a stationary state $\psi(x,t) = \ee^{-\ii Et}\phi(x)$ it is conserved, since $|\psi|$ and hence $Q$ are time-independent. However, we also have that ${\mathcal E}$ takes the same value for all possible trajectories, namely ${\mathcal E} = E$. Hence this is a trivial constant of motion and it does not reduce the effective number of degrees of freedom. Therefore the Poincar\'e-Bendixson theorem does not exclude the possibility of chaos. In section \ref{1particle}, we will provide some examples of stationary states in 3-d that lead to chaos.

\subsection{Participation ratio and measures of entanglement}\label{measures}

In this section, we present simple quantities which we will compare to the average Lyapunov exponent later on.

First, given a decomposition $\psi=\sum_{i=1}^N c_i \phi_i$ in an orthonormal basis $\phi_i$, $i=1,\dots,N$, the participation ratio (PR)~\cite{Viola07} is defined by
\begin{equation}
\textrm{PR}(\psi)=\frac{1}{\sum_{i=1}^N |\langle\phi_i|\psi\rangle|^4}=\frac{1}{\sum_{i=1}^N |c_i|^4}.
\label{PR}
\end{equation}
It quantifies the number of basis states on which the state $\psi$ is delocalized. The PR is equal to $1$ when $\psi$ is a basis element $\phi_i$ and takes the maximum value $N$ when $\psi$ is an equally weighted superposition of all the basis states. 

For many-particle systems, entanglement will couple the dynamics of the different particles. So the entanglement will generically play a role in the generation of chaos. Here, we will consider three measures of entanglement: the Meyer-Wallach measure ($Q$), the geometric measure ($E_G$) and the three-tangle ($\tau_3$). For 3-qubit states, the Meyer-Wallach measure of entanglement $Q(\psi) \in [0,1]$ is defined by~\cite{meyer02}
\be
Q(\psi)=\frac{1}{3}\:\sum_{k=1}^{3}2(1-\mathrm{tr}{\widehat \rho}_{k}^2),
\label{Q}
\en
with ${\widehat \rho}_{k}$ the reduced density matrix for the $k$-th qubit. It is the average of the linear entropies $2(1-\mathrm{tr}{\widehat \rho}^2_k)$ of each qubit~\cite{meyer02}. 

The geometric measure of entanglement $ E_G(\psi) \in [0,1[$ is defined as~\cite{wei03}:
\be
E_G(\psi)=1-\max_{|\phi\rangle~\mathrm{sep.}}|\langle\phi|\psi\rangle|^2,
\label{E_G}
\en
where the maximum is taken over all possible separable (i.e.\ non-entangled) states. It can be interpreted as the distance of a state $|\psi\rangle$ to the set of separable states. 

For a three-qubit state 
\be
|\psi\rangle=\sum_{i,j,k=0,1}c_{ijk}|ijk\rangle,
\en
with $|ijk\rangle$ the basis states, the three-tangle is defined as~\cite{coff00}:
\begin{equation}
\tau_3(\psi)=4|d_1-2d_2+4d_3|,
\label{tau_3}
\end{equation}
with
\begin{align*}
d_1{}={}&c_{000}^2c_{111}^2+c_{001}^2c_{110}^2+c_{010}^2c_{101}^2+c_{100}^2c_{011}^2,\\[3pt]
d_2{}={}&c_{000}c_{111}c_{011}c_{100}+c_{000}c_{111}c_{101}c_{010}+c_{000}c_{111}c_{110}c_{001}\\
& +c_{011}c_{100}c_{101}c_{010}+c_{011}c_{100}c_{110}c_{001}+c_{101}c_{010}c_{110}c_{001},\\[3pt]
d_3{}={}&c_{000}c_{110}c_{101}c_{011}+c_{111}c_{001}c_{010}c_{100}.
\end{align*}
It can be interpreted as the amount of entanglement between one qubit and the remaining two qubits not accounted for by the amount of entanglement between pairs of qubits. All these entanglement measures vanish if the state $\psi$ is separable. Conversely, when the measure vanishes this implies that the state $\psi$ is separable in the case of the first two measures, but not in the case of the three-tangle.

\section{States with moving nodes and no chaos}\label{nochaos}
As we will show in the next section, one can have wave functions with stationary nodes that give rise to chaotic Bohmian trajectories. Conversely, having (arbitrarily many) moving nodes is no guarantee for chaos, as we will illustrate here with some examples. A more complex mechanism is required, as for example discussed in \cite{efthymiopoulos07,efthymiopoulos09} for two dimensional systems.

In one spatial dimension, a non-stationary state will generically have moving nodes, but the Bohmian motion will not be chaotic due to the Poincar\'e-Bendixson theorem.

For two or three dimensions, consider the superposition of two stationary states of the form
\begin{equation}
\psi(\mathbf{x},t)=c_1 \ee^{-\ii E_1t }\phi_1(\mathbf{x})+c_2 \ee^{-\ii E_2t }\phi_2(\mathbf{x}),
\label{1part_2st}
\end{equation}
where the $c_i$, $i=1,2$, are complex numbers and where the $\phi_i$ are of the form (\ref{5}), i.e.\ a real function that is separable in the Cartesian coordinates. Such states generically have moving nodes (considering states $\phi_i$ with different energies). However, there are two constants of motion of the type (\ref{com}) in the case of three dimensions and one in the case of two dimensions. These constants of motion reduce the effective number of degrees of freedom so that, again as a consequence of the Poincar\'e-Bendixson theorem, the Bohmian motion can not be chaotic. 

For the final example, consider again superpositions of the form~(\ref{1part_2st}), but now with the $\phi_i$ of the form (\ref{6}), i.e., $\phi_{i}(r,\theta,\varphi)=f_{i}(r,\theta)\ee^{\ii m_i \varphi}$, with the $f_i$ real. The number of nodes of the functions $f_i$ grows with the energy. As such, we could construct states with an arbitrary number of moving nodes. These nodes move in circles around the $z$-axis with a constant angular velocity. This is because the state becomes stationary in a rotating frame, with angular frequency
\begin{equation}
\omega=\frac{E_1-E_2}{m_2-m_1}.
\end{equation}
Namely, if we introduce the variable $\varphi'=\varphi + \omega t$, then the state (\ref{1part_2st}) reads 
\be
\psi(r,\theta,\varphi)= \psi'(r,\theta,\varphi',t) = \ee^{- \ii(E_1 +m_1 \omega)t +\ii m_1 \varphi' }\left( c_1 f_1(r,\theta)+c_2 f_2(r,\theta) \ee^{\ii \varphi'(m_2 - m_1)} \right) .
\en 
So in the rotating frame the state is stationary and the Bohmian motion reduces to the autonomous dynamics
\be
\frac{dr}{dt} = \frac{\pa S'}{\pa r} \,, \qquad \frac{d\theta}{dt} = \frac{1}{r^2} \frac{\pa S'}{\pa \theta} \,, \qquad  \frac{d\varphi'}{dt} = \frac{1}{r^2 \sin^2\theta}\frac{\pa S'}{\pa \varphi'} + \omega,
\label{7a}
\en
where $S'$ is the phase of $\psi'$. In two dimensions this implies that the Bohmian motion can not be chaotic because of the Poincar\'e-Bendixson theorem (as was noted before in \cite{makowski00}). The same conclusion holds for three dimensions since in this case there is also a constant of motion of the form \eqref{com2} for $r$ and $\theta$.

\section{Single particle in 3-d}\label{1particle}
We have already mentioned that for a stationary state in 2-d, the motion can not be chaotic. Therefore we need to consider at least three dimensions. We will consider stationary superpositions $\psi = \ee^{-\ii Et}  \phi =  \ee^{-\ii Et} \sum_i \phi_i$ of three and four orthogonal degenerate energy eigenstates $\phi_i$ of the form (\ref{5}) or (\ref{6}). As shown in the previous section, there is no chaos for superpositions of only two such states.

\subsection{Superposition of three eigenstates: examples of chaotic motion}
For the harmonic oscillator we can consider the complete set of energy eigenstates $\phi^{\textrm{3d}}_{n_x,n_y,n_z}(x,y,z)$, cf.~(\ref{eigenstate_har_3dR}), or $\phi^{\textrm{sph}}_{k,l,m}(r,\theta,\varphi)$, cf.~(\ref{eigenstate_har_3dsph}). These states are respectively of the form (\ref{5}) and (\ref{6}). For each choice, we will consider a superposition of three states that gives rise to chaotic Bohmian motion. 

As a first example, consider the superposition
\be
\phi^{\textrm{sph}}(r,\theta,\varphi)=\phi^{\textrm{sph}}_{0,3,1}(r,\theta,\varphi)+\ee^{\ii\pi/3}\phi^{\textrm{sph}}_{0,3,0}(r,\theta,\varphi)+\ee^{\ii\pi/7}\phi^{\textrm{sph}}_{1,1,0}(r,\theta,\varphi)
\label{ho3d_stat}
\en
with energy $9/2$. This state has nodal lines, which are however difficult to find analytically.
\begin{figure}[htb]
\centering
\includegraphics{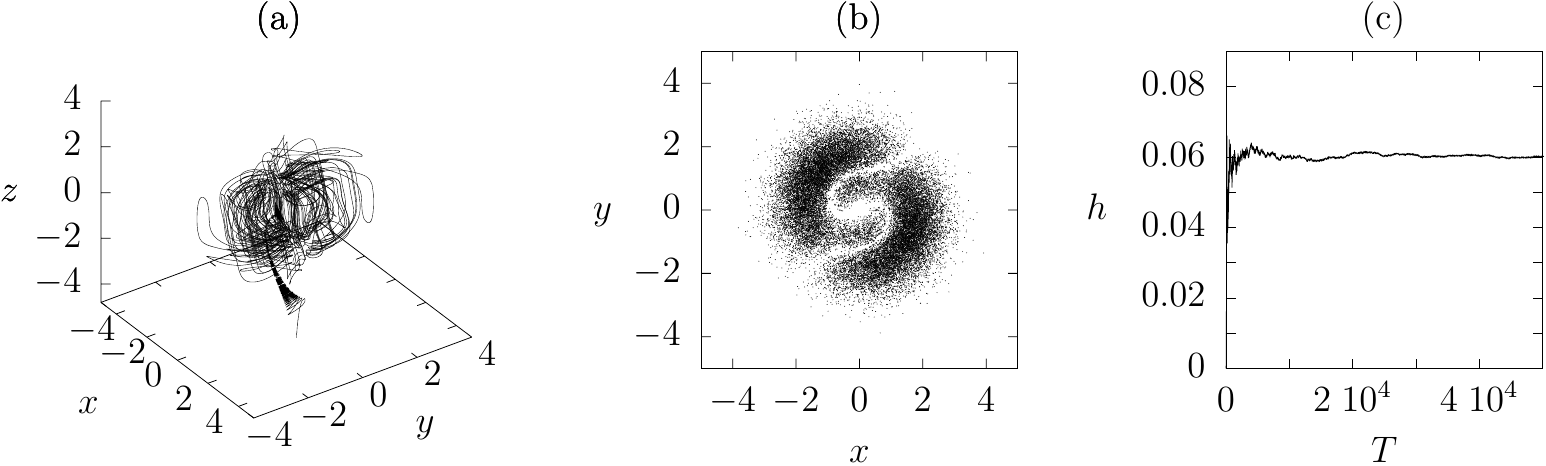}
\caption{(a) Bohmian trajectory for a particle with wave function \eqref{ho3d_stat} and initial position $(r,\theta,\varphi)=(6.6969,2.38696,-0.249865)$. (b) Poincar\'e section in the plane $z=0$. (c) Evolution of $h(\mathbf{x}_0,T)$ as a function of time for the trajectory depicted in $(a)$.}
\label{HO3d_stat_sph}
\end{figure}
Figure~\ref{HO3d_stat_sph} illustrates a typical Bohmian trajectory for this wave function.  The trajectory starting outside the bulk of the wave packet moves towards it and eventually moves inside it. The Poincar\'e section in the plane $z=0$ shown on the middle panel of figure~\ref{HO3d_stat_sph} indicates that the dynamics is not regular. This is confirmed by the Lyapunov exponent $h\approx 0.06$ (see the right panel of figure~\ref{HO3d_stat_sph}).

As a second example, consider now the stationary state 
\be
\phi^{\textrm{3d}}(x,y,z)=\phi^{\textrm{3d}}_{1,1,1}(x,y,z)+\ee^{\ii\pi/3}\phi^{\textrm{3d}}_{3,0,0}(x,y,z)+\ee^{\ii\pi/7}\phi^{\textrm{3d}}_{1,2,0}(x,y,z),
\label{HO3d_stat_cart}
\en
which also has energy $9/2$. In this case there is a nodal plane at $x=0$. In the limit of $x$ going to zero, the velocity does not blow up but rather becomes tangential to the plane $x=0$. The Bohmian trajectories do not cross this plane, they tend to be repelled by it. The other nodal lines are 
given by
\begin{multline}
\Bigg\lbrace(x,y,z): x=\pm\sqrt{\frac{3}{2}+\sin\left(\frac{\pi}{7}\right)-2y^2\sin\left(\frac{\pi}{7}\right)},
  y\neq 0, \\
  |y|\leq \frac{1}{2}\sqrt{\frac{3+2\sin(\pi/7)}{\sin(\pi/7)}}, 
  z=\frac{(1-2y^2)[3\cos(\pi/7)-\sqrt{3}\sin(\pi/7)]}{6\sqrt{2}y} \Bigg\rbrace.
\end{multline}
These nodal lines end in the plane $x=0$.
\begin{figure}[htb]
\centering
\includegraphics{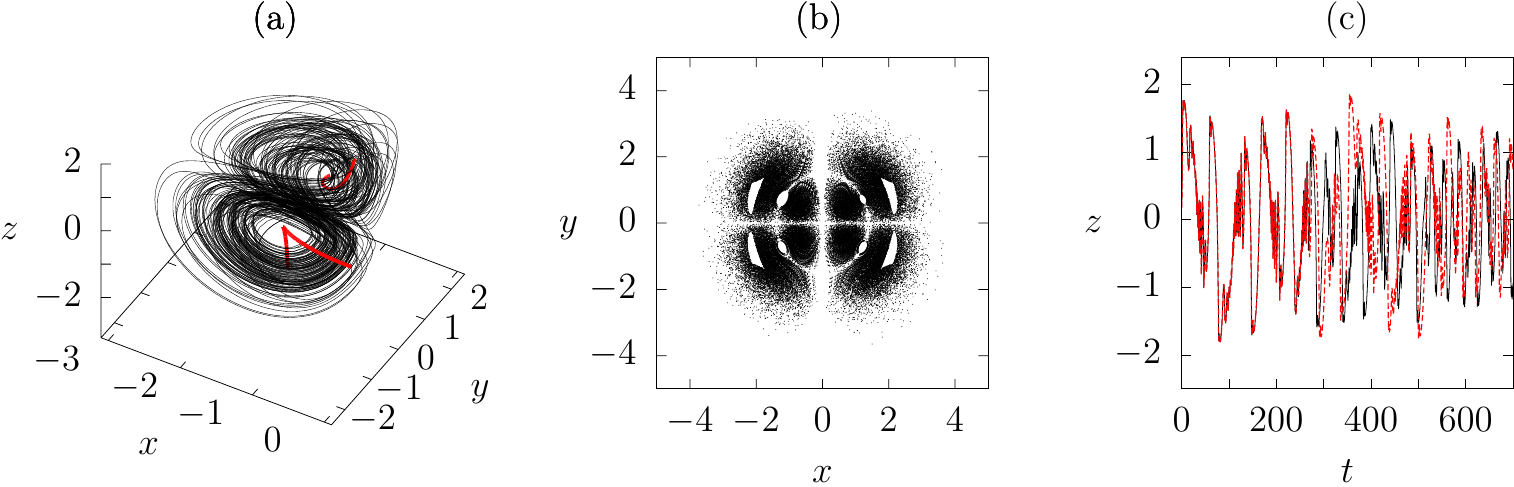}
\caption{(a) Bohmian trajectory for a particle with wave function \eqref{HO3d_stat_cart} and initial position $\mathbf{x}_0\equiv(x,y,z)=(-2.212756,-1.97466,0.179963)$. It spirals around the nodal lines, which are plotted in red (gray in grayscale) (b) Poincar\'e section in the plane $z=0$. (c) $z$-component of the trajectories starting at $\mathbf{x}_0$ (black) and $\mathbf{x}_0+(0,10^{-6},0)$ (red dashed). The divergence is exponential. However, since the Lyapunov exponent is 0.06, this becomes visible to the naked eye only at around $t=300$.}
\label{HO3d_stat_rect}
\end{figure}
A typical Bohmian trajectory for this system is illustrated in figure~\ref{HO3d_stat_rect}. The trajectory first spirals around a nodal line. When the particle arrive at the end of the nodal line, the trajectory continues freely until it reaches the vicinity of another nodal line. Such behavior has been reported before~\cite{frisk97,falsaperla03}. Most of the Poincar\'e section in the plane $z=0$ is composed of regions where the points look randomly distributed which constitutes a strong clue that the motion is chaotic. This is confirmed by the calculated Lyapunov exponent which is again approximately equal to $0.06$. The right panel of figure~\ref{HO3d_stat_rect} illustrates the exponential divergence of two initially close trajectories.

\subsection{Chaos and participation ratio}\label{sup4modes}

We have shown that the Bohmian trajectories of 3-d  systems with stationary states may be chaotic. We now study the effect of the particular form of the wave function on the amount of chaos. We will consider the harmonic oscillator and stationary wave functions of the form 
\begin{equation}
\phi(\mathbf{x};\alpha,\beta)=\sum^4_{i=1}{c_i(\alpha,\beta)\phi^{\textrm{sph}}_{k_i,l_i,m_i}(\mathbf{x})}, 
\label{psi_alpha}
\end{equation}
where $c_i(\alpha,\beta)=|c_i(\alpha)|\ee^{\ii \chi_i(\beta)}$, so that the amplitude and the phase of the coefficients $c_i$ are functions of respectively $\alpha$ and $\beta$. The quantum numbers $k_i,l_i,m_i$ are chosen such that $\phi(\mathbf{x};\alpha,\beta)$ is a stationary state. We also assume the states to be normalized (i.e., $\sum_i |c_i|^2 = 1$). While this does not affect the Bohmian velocity field, it will be important when considering the participation ratio. In order to quantify the amount of chaos, we compute the average Lyapunov exponent ${\bar h}$ over $150$ different initial positions, uniformly distributed in a cube of edge length $10$ centred around the origin. The trajectories used in the determination of the Lyapunov exponent were computed from $t=0$ to $t=5 \cdot 10^4$.

Consider first the states
\begin{multline}
{\phi}(\mathbf{x};\alpha,\beta)= \mathcal{N}(\alpha)[\cos(\alpha)\phi^{\textrm{sph}}_{1,3,0}(\mathbf{x}) +\ee^{\ii (\beta+\pi/3)}\sin(\alpha)\phi^{\textrm{sph}}_{1,3,1}(\mathbf{x})\\ 
 +\ee^{\ii(2\pi \cos(\beta)+\pi/5)}\cos^2(\alpha)\phi^{\textrm{sph}}_{2,1,-1}(\mathbf{x}) + \ee^{\ii (-2\beta+ \pi/7)}\sin^2(\alpha)\phi^{\textrm{sph}}_{2,1,0}(\mathbf{x})],
\label{31}
\end{multline}
with $\alpha \in [0,\pi/2]$, $\beta \in [0,2\pi]$ and $\mathcal{N}(\alpha)$ a normalization constant. The energy of these states is $13/2$.

\begin{figure}[t]
\centering
\includegraphics{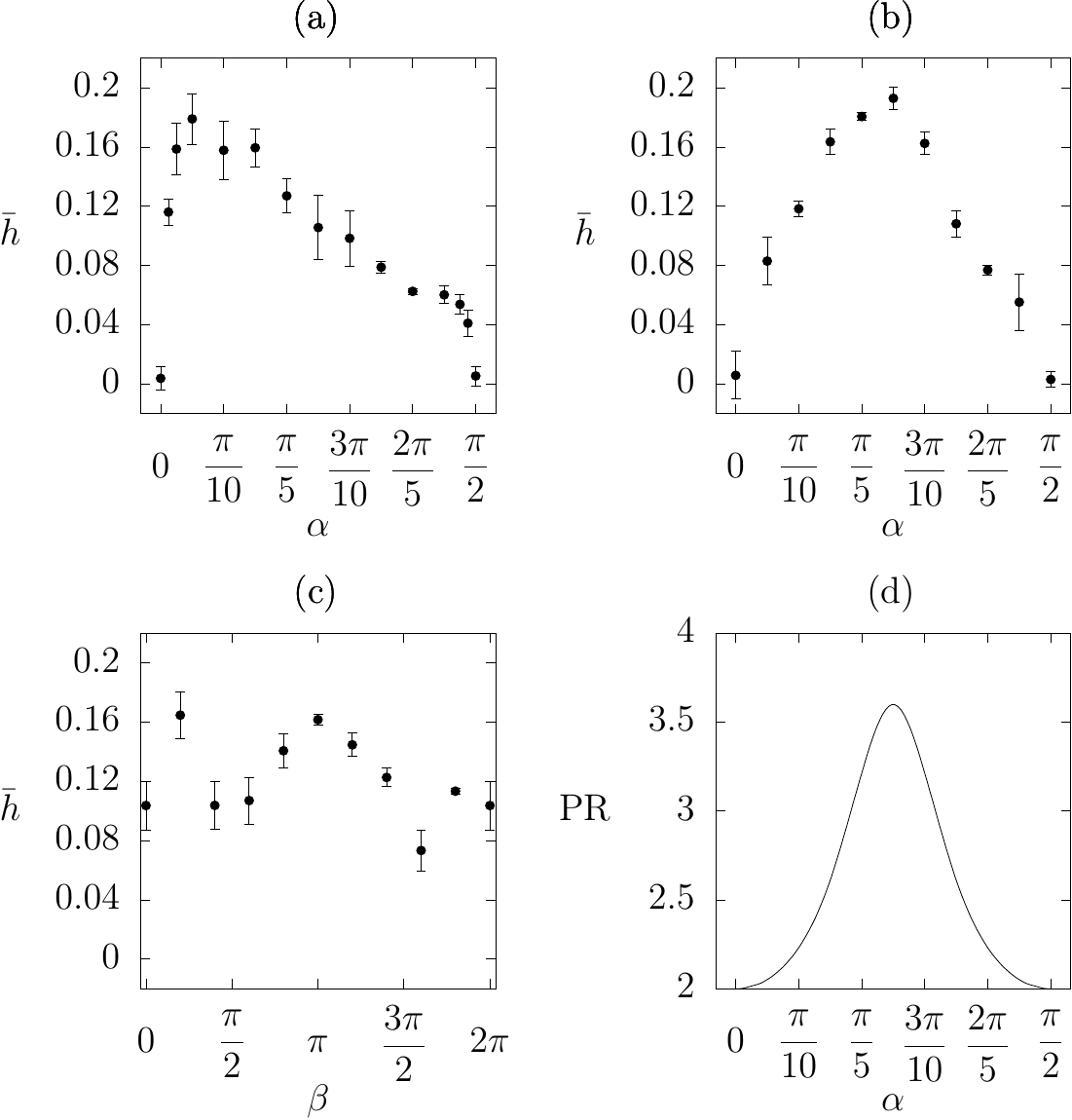}
\caption{(a) and (b) Average Lyapunov exponent $\bar{h}$ as a function of $\alpha$ for Bohmian trajectories for respectively the states ${\phi}(\mathbf{x};\alpha,0)$ of energy $13/2$ and ${\phi'}(\mathbf{x};\alpha,0)$ of energy $17/2$. The average $\bar{h}$ is performed over $150$ initial positions. The error bars correspond to the standard deviation of the Lyapunov exponent over different trajectories. (c) $\bar{h}$ as a function of $\beta$ for the state ${\phi}(\mathbf{x};\pi/4,\beta)$. (d) Participation ratio as a function of $\alpha$ for the states ${\phi}(\mathbf{x};\alpha,0)$ and ${\phi'}(\mathbf{x};\alpha,0)$. }
\label{hbarandpr}
\end{figure}

Taking $\beta=0$, so that the phases of the coefficients are fixed, we have
\begin{multline}
{\phi}(\mathbf{x};\alpha,0)= \mathcal{N}(\alpha)[\cos(\alpha)\phi^{\textrm{sph}}_{1,3,0}(\mathbf{x}) +\ee^{\ii \pi/3}\sin(\alpha)\phi^{\textrm{sph}}_{1,3,1}(\mathbf{x})\\ 
 +\ee^{\ii\pi/5}\cos^2(\alpha)\phi^{\textrm{sph}}_{2,1,-1}(\mathbf{x}) + \ee^{\ii \pi/7}\sin^2(\alpha)\phi^{\textrm{sph}}_{2,1,0}(\mathbf{x})].
\end{multline}
Figure~\ref{hbarandpr}(a) shows the average Lyapunov exponent as a function of $\alpha$.
For $\alpha=0$ and $\alpha=\pi/2$, ${\phi}(\mathbf{x};\alpha,0)$ is a superposition of only two eigenstates and thus the Bohmian trajectories are regular, cf.\ section \ref{cst_of_motion}. This agrees with the calculated value of $\bar{h}$ which is zero. $\bar{h}$ is maximum around $\alpha=\pi/10$ and then decreases as $\alpha$ increases.

In order to investigate the effect of the choice of eigenstates in ${\phi}(\mathbf{x};\alpha,\beta)$ on the chaotic behavior of the Bohmian trajectories, we also compute the average Lyapunov exponent for the stationary state
\begin{multline}
{\phi'}(\mathbf{x};\alpha,0)= \mathcal{N}(\alpha)[\cos(\alpha)\phi^{\textrm{sph}}_{2,3,2}(\mathbf{x}) +\ee^{\ii \pi/3}\sin(\alpha)\phi^{\textrm{sph}}_{2,3,-1}(\mathbf{x})\\ 
 +\ee^{\ii \pi/5}\cos^2(\alpha)\phi^{\textrm{sph}}_{3,1,1}(\mathbf{x}) + \ee^{\ii  \pi/7}\sin^2(\alpha)\phi^{\textrm{sph}}_{3,1,0}(\mathbf{x})],
\label{32}
\end{multline}
which has energy $17/2$. This state differs from  ${\phi}(\mathbf{x};\alpha,0)$ only in the quantum numbers of the four eigenstates in terms of which it is constructed. Figure~\ref{hbarandpr}(b) shows the average Lyapunov exponent for some values of $\alpha$. Just as before, for $\alpha=0$ and $\alpha= \pi/2$, $\bar{h}$ should vanish. For ${\phi'}(\mathbf{x};\alpha,0)$, $\bar{h}$ reaches the maximal value approximately equal to $0.2$ near $\alpha=\pi/4$.

Although the states ${\phi}(\mathbf{x};\alpha,0)$ and ${\phi'}(\mathbf{x};\alpha,0)$ have the same coefficients in their decomposition in the basis of eigenstates of the harmonic oscillator, the evolution of the average Lyapunov exponent is significantly different. For both wave functions, $\bar{h}$ takes a maximum value of around $0.2$ but for a different value of $\alpha$. So, in particular, a small increase of the energy of the wave function does not necessarily lead to a greater amount of chaos in the Bohmian trajectories. For instance, one can see by comparing figures~\ref{hbarandpr}(a) and~\ref{hbarandpr}(b) that for $\alpha=\pi/10$, $\bar{h}$ is more than twice as high for ${\phi}(\mathbf{x};\alpha)$ as for ${\phi'}(\mathbf{x};\alpha,0)$, even though the energy of the latter wave function is higher. When $\alpha=\pi/4$ this situation is reversed.

In order to bring to light the effect of the choice of phases of the coefficients, we consider the states ${\phi}(\mathbf{x};\pi/4,\beta)$, where the amplitudes of the coefficients are fixed and the phases may vary. Figure~\ref{hbarandpr}(c) illustrates the average Lyapunov exponent for some values of the parameter $\beta$. The average Lyapunov exponent takes values between $0.07$ and $0.16$. Thus, $\bar{h}$ is quite sensitive to the choice of the relative phase of the coefficients. The dependence on $\beta$ also looks rather complicated. 

Let us now compare the average Lyapunov exponent with the participation ratio (PR)~\eqref{PR}. It is immediately clear that the PR has two shortcomings, which make that it can not agree perfectly with the average Lyapunov exponent. First, the PR does not depend on the particular form of the basis states. Therefore, it takes the same value for the wave functions ${\phi}(\mathbf{x};\alpha,0)$ and ${\phi'}(\mathbf{x};\alpha,0)$ (cf.\ \ref{31} and \ref{32}). Figure~\ref{hbarandpr}(d) shows the PR for these states as a function of $\alpha$. The PR shows a good qualitative agreement with the average Lyapunov exponent $\bar{h}$ for the wave function ${\phi'}(\mathbf{x};\alpha,0)$, but a bit less so for ${\phi}(\mathbf{x};\alpha,0)$. For both ${\phi}(\mathbf{x};\alpha,0)$ and ${\phi'}(\mathbf{x};\alpha,0)$, the PR reaches a minimum for $\alpha=0$ and $\alpha= \pi/2$, just as $\bar{h}$ (whose theoretical value is zero). The PR has a maximum at $\alpha=\pi/4$ as does the average Lyapunov exponent for the wave function ${\phi'}(\mathbf{x};\alpha,0)$. However, for the wave function ${\phi}(\mathbf{x};\alpha,0)$ the maxima of $\bar{h}$ and $\textrm{PR}$ do not coincide. And while the PR is symmetric around $\alpha=\pi/4$, like $\bar{h}$ for ${\phi'}(\mathbf{x};\alpha,0)$, $\bar{h}$ for ${\phi}(\mathbf{x};\alpha,0)$ is not.

The second shortcoming of the PR is that it does not depend on the phases of the coefficients $c_i$. As we have seen, with the wave functions ${\phi}(\mathbf{x};\pi/4,\beta)$, the amount of chaos strongly depends on the value of the phases, cf.\ figure~\ref{hbarandpr}(c). 

Therefore, in order to characterize the amount of chaos in Bohmian trajectories, a simple measure of superposition such as the PR is not enough. One needs to take into account the particular form of the eigenstates that appear in the superposition as well as their relative phases, since both influence the complexity of the wave function. Nevertheless, generically, the PR gives the general trend of the value of the Lyapunov exponent.

\section{Two-particle systems}\label{2particle}
In this section we study the Bohmian trajectories of systems of two particles whose wave functions are stationary entangled states. We focus our attention on 2-d  systems. Our interest lies of course with entangled states, since for such states the motion of one particle depends on the position of the other particle, unlike for a separable state. We give three examples of wave functions for which there is chaotic motion. 

\subsection{Superpositions with 2 single-particle basis states}
Using two single-particle basis states $\phi_1$ and $\phi_2$, the most general two-particle state reads
\begin{equation}
\phi^{2\mathrm{p}}_{2}(\mathbf{x}_1,\mathbf{x}_2)=c_1\phi_1(\mathbf{x}_1)\phi_{1}(\mathbf{x}_2)+c_2\phi_{1}(\mathbf{x}_1)\phi_{2}(\mathbf{x}_2)
+c_3\phi_{2}(\mathbf{x}_1)\phi_{1}(\mathbf{x}_2)+c_4\phi_{2}(\mathbf{x}_1)\phi_{2}(\mathbf{x}_2)
\label{2st_2part}
\end{equation}
where the superscript denotes the number of particles and the subscript the number of single-particle basis states involved in the superposition. For states $\phi_i(x,y)$ of the form~(\ref{5}) there are two constants of motion of the form \eqref{com} (even for a non-stationary superposition){\footnote{This follows from the fact that one can write $\phi^{2\mathrm{p}}_{2}$ in two different ways in the form \eqref{8}, namely
\begin{equation}
\phi^{2\mathrm{p}}_{2}(\mathbf{x}_1,\mathbf{x}_2) = \phi_1(\mathbf{x}_1)\left( c_1 \phi_{1}(\mathbf{x}_2)+c_2\phi_{2}(\mathbf{x}_2) \right) + \phi_{2}(\mathbf{x}_1) \left(c_3\phi_{1}(\mathbf{x}_2)+c_4\phi_{2}(\mathbf{x}_2)\right)
\end{equation}
and
\begin{equation}
\phi^{2\mathrm{p}}_{2}(\mathbf{x}_1,\mathbf{x}_2) = \phi_{1}(\mathbf{x}_2) \left(c_1\phi_1(\mathbf{x}_1)+c_3\phi_{2}(\mathbf{x}_1) \right)+ \phi_{2}(\mathbf{x}_2) \left(c_2\phi_{1}(\mathbf{x}_1) +c_4\phi_{2}(\mathbf{x}_1) \right).
\end{equation}
Since the $\phi_i(x,y)$ are of the form (\ref{5}), there are two constants of motion of the form \eqref{com} for $(x_1,y_1)$ and $(x_2,y_2)$.}}, so that the effective number of degrees of freedom is 2 and no chaos is possible. 

Consider states of the form (\ref{2st_2part}), but now with the $\phi_i(r,\varphi)$ of the form~(\ref{6})). Unlike the case where they are of the form  \eqref{5}, such states generically do not seem to yield constants of motion, so that chaotic motion may be possible. As an example, consider the harmonic oscillator with single-particle energy eigenstates $\phi^{\textrm{pol}}_{n_r,n_l}(r,\varphi)$, cf.~(\ref{eigenstate_har_2dC}), and the wave function
\begin{multline}
\phi^{2\mathrm{p}}_{2}(\mathbf{x}_1,\mathbf{x}_2)=\phi^{\textrm{pol}}_{1,1}(\mathbf{x}_1)\phi^{\textrm{pol}}_{1,1}(\mathbf{x}_2)+\ee^{\ii\pi/3}\phi^{\textrm{pol}}_{1,1}(\mathbf{x}_1)\phi^{\textrm{pol}}_{2,0}(\mathbf{x}_2)\\
+\ee^{\ii\pi/5}\phi^{\textrm{pol}}_{2,0}(\mathbf{x}_1)\phi^{\textrm{pol}}_{1,1}(\mathbf{x}_2)+\ee^{\ii\pi/7}\phi^{\textrm{pol}}_{2,0}(\mathbf{x}_1)\phi^{\textrm{pol}}_{2,0}(\mathbf{x}_2).
\label{60}
\end{multline}
\begin{figure}[t]
\centering
\includegraphics{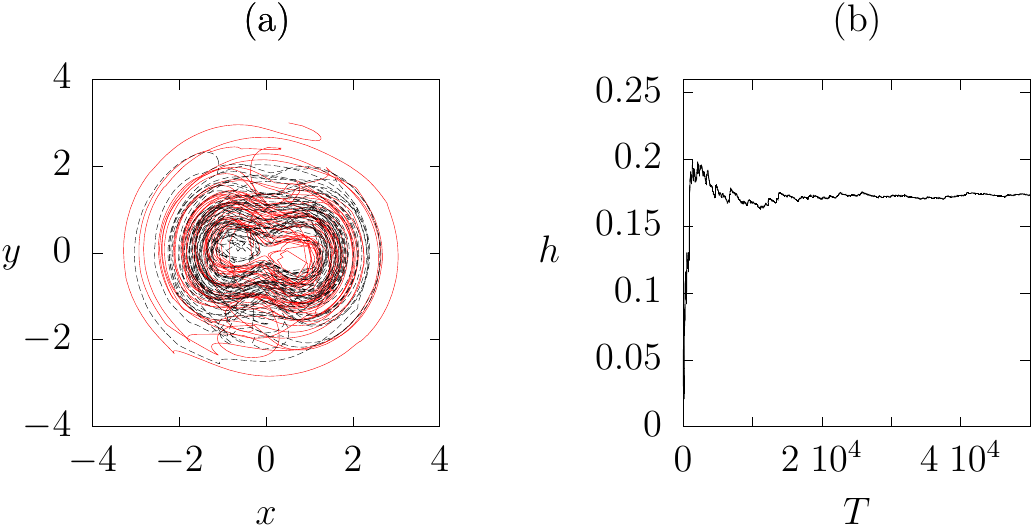}
\caption{(a) Bohmian trajectories for a system of two particles for the entangled wave function (\ref{60}) and initial configuration $(x_1,y_1,x_2,y_2)=(2.37166,-0.374916,-0.522219,2.99893)$. The black dashed curve corresponds to the trajectory of particle $1$ and the red curve to that of particle $2$. (b) Evolution of $h(\mathbf{x}_0,T)$ as a function of time for the trajectory depicted in $(a)$.}
\label{2dharmonic_2part_2st}
\end{figure}
Figure~\ref{2dharmonic_2part_2st} illustrates a typical set of trajectories for this system. The Lyapunov exponent takes a value $h\approx 0.17$, which indicates chaos.

\subsection{Superpositions with 3 single-particle basis states}
As we have seen in the previous section, stationary superpositions of only two single-particle basis states  $\phi_i(x,y)$  of the form~\eqref{5} cannot give rise to chaos. For superpositions using three such states, we show that chaos is possible. We give 2 examples for states of the form 
\begin{equation}
\phi^{2\mathrm{p}}_{3}(\mathbf{x}_1,\mathbf{x}_2)=c_1\phi_{1}(\mathbf{x}_1)\phi_{1}(\mathbf{x}_2)+c_2\phi_{2}(\mathbf{x}_1)\phi_{2}(\mathbf{x}_2)+c_3\phi_{3}(\mathbf{x}_1)\phi_{3}(\mathbf{x}_2),
\end{equation}
one for the harmonic oscillator and one for the square box. Generically such states do not yield constants of motion.

As a first example, we consider the state
\be
\phi^{2\mathrm{p}}_{3} = \phi^{\textrm{2d}}_{1,1}\phi^{\textrm{2d}}_{1,1}+\ee^{\ii\pi/3}\phi^{\textrm{2d}}_{2,0}\phi^{\textrm{2d}}_{2,0}+\ee^{\ii\pi/7}\phi^{\textrm{2d}}_{0,2}\phi^{\textrm{2d}}_{0,2}
\label{50}
\en
for the 2d harmonic oscillator with the energy eigenstates $\phi^{\textrm{2d}}_{n_x,n_y}$ given in~(\ref{eigenstate_har_2dR}). 
\begin{figure}[htb]
\centering
\includegraphics{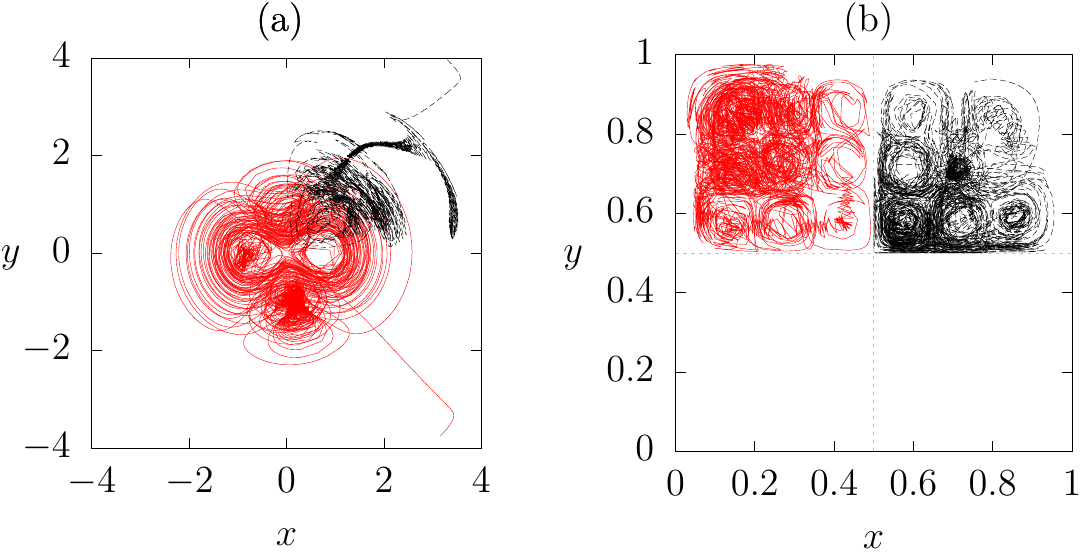}
\caption{(a) and (b) Bohmian trajectories for a system of two particles respectively for the entangled wave functions (\ref{50}) and (\ref{52}), and with initial configuration respectively $(x_1,y_1,x_2,y_2)=(3.29867,3.97517,3.15679,-3.75662)$ and $(x_1,y_1,x_2,y_2)=(0.666891,0.584026,0.193745,0.747208)$. The black dashed curve corresponds to the trajectory of particle $1$ and the red curve to that of particle $2$.}
\label{2dharmonic_2part_3st}
\end{figure}
Figure~\ref{2dharmonic_2part_3st}(a) illustrates a typical pair of Bohmian trajectories for this system. Again, the trajectories start outside the bulk of the wave packet move towards it and eventually move inside it. The Lyapunov exponent is $h\approx 0.08$, so that the motion is chaotic.

As a second example, we consider the state
\be
\phi^{2\mathrm{p}}_{3} = \phi^{\textrm{box}}_{7,1}\phi^{\textrm{box}}_{7,1}+  \ee^{\ii\pi/3}\phi^{\textrm{box}}_{1,7}\phi^{\textrm{box}}_{1,7} +\ee^{\ii\pi/7}\phi^{\textrm{box}}_{5,5}\phi^{\textrm{box}}_{5,5}
\label{52}
\en
for the 2d square box with the eigenstates $\phi^{\textrm{box}}_{n_x,n_y}$ given in~(\ref{eigenstate_box_2d}). 
Figure~\ref{2dharmonic_2part_3st}(b) shows a typical set of Bohmian trajectory for this system. We obtain the rather large value of around $25$ for the Lyapunov exponent. This example also illustrates the non-local character of Bohmian mechanics. Although the particles are always in different regions of the box, their trajectories strongly influence each other which allows chaotic motion (remember that for a 2-d system  with stationary wave function, the Bohmian trajectories are always regular). Actually, the amount of chaos would not change if we moved one particle in a box arbitrarily far away.

\section{Three-particle systems}\label{3particle}
In this section, we consider 3-particle systems in 2-d and investigate the amount of chaos for different entangled states.

\subsection{Superpositions with 2 single-particle basis states}\label{2bs}

First we consider superpositions formed using two different single-particle states $\phi_i$. These states can be seen as three-qubit states. We focus our attention on the states
\begin{equation}
\phi^{3\mathrm{p}}_{\textrm{GHZ}}(\mathbf{x}_1,\mathbf{x}_2,\mathbf{x}_3)=c_1 \phi_1(\mathbf{x}_1)\phi_1(\mathbf{x}_2)\phi_1(\mathbf{x}_3)+c_2 \phi_2(\mathbf{x}_1)\phi_2(\mathbf{x}_2)\phi_2(\mathbf{x}_3),
\label{3part_GHZ}
\end{equation}
\begin{equation}
\phi^{3\mathrm{p}}_{\textrm{W}}(\mathbf{x}_1,\mathbf{x}_2,\mathbf{x}_3)=c_1 \phi_1(\mathbf{x}_1)\phi_2(\mathbf{x}_2)\phi_2(\mathbf{x}_3)+c_2 \phi_2(\mathbf{x}_1)\phi_1(\mathbf{x}_2)\phi_2(\mathbf{x}_3)+c_3 \phi_2(\mathbf{x}_1)\phi_2(\mathbf{x}_2)\phi_1(\mathbf{x}_3),
\label{3part_W}
\end{equation}
which belong respectively to the Greenberger-Horne-Zeilinger (GHZ) and the W entanglement classes~\cite{duer00}.

For states $\phi_i(x,y)$ of the form (\ref{5}), there are 5 independent constants of motion of the form \eqref{com}, for $\phi^{3\mathrm{p}}_{\textrm{GHZ}}$. The states  $\phi^{3\mathrm{p}}_{\textrm{W}}$ lead to 3 constants of motion of the form \eqref{com} for $(x_1,y_1)$, $(x_2,y_2)$ and $(x_3,y_3)$, and 2 of the form \eqref{com3} for $(x_1,x_2,x_3)$ and $(y_1,y_2,y_3)$. In both cases, the constants of motion make that there is no chaos, despite the entanglement. This remains true even if the states are non-stationary (and thus have moving nodes).

For states $\phi_i(r,\varphi)$ of the form (\ref{6}), the state $\phi^{3\mathrm{p}}_{\textrm{GHZ}}$ gives two independent constants of motion of the form \eqref{com} for $(r_1,r_2)$ and $(r_2,r_3)$. If we introduce the variable $\varphi = \varphi_1 + \varphi_2 + \varphi_3$, then the wave function depends on this variable and $r_1$, $r_2$ and $r_3$. So in terms of the variables $\varphi,\varphi_2,\varphi_3,r_1,r_2,r_3$, the Bohmian dynamics for $\varphi,r_1,r_2,r_3$ is independent of $\varphi_2$ and $\varphi_3$. The two constants of motion imply that the dynamics for $\varphi,r_1,r_2,r_3$ is regular. Furthermore, the dynamics of $\varphi_2$ and $\varphi_3$ only depends on $\varphi,r_1,r_2,r_3$, and therefore there is no chaos.

\begin{figure}[ht]
\centering
\includegraphics{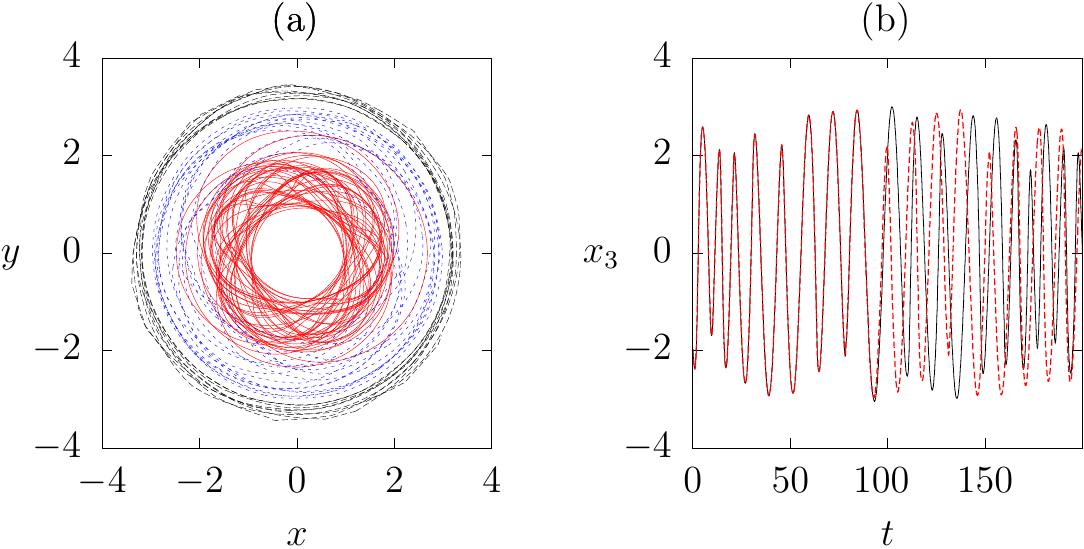}
\caption{Bohmian trajectories for a system of three particles in a $2$-d  harmonic potential for the entangled  wave function (\ref{80}) and initial configuration $\mathbf{x}_0\equiv(x_1,y_1,x_2,y_2,x_3,y_3)=(1.40802, -3.0515, 0.97766, 1.33025, -1.971814, 1.64945)$. The black dashed curve corresponds to the trajectory of particle $1$, the red curve to that of particle $2$ and the blue one to that of particle $3$. (b) $x_3$-component of the trajectories starting at $\mathbf{x}_0$ (black) and $\mathbf{x}_0+(0,10^{-6},10^{-6},0,0,0)/\sqrt{2}$ (red dashed).}
\label{2dharmonic_3part}
\end{figure}

For states $\phi_i(r,\varphi)$ of the form (\ref{6}), the state $\phi^{3\mathrm{p}}_{\textrm{W}}$ leads to one constant of motion of the type~\eqref{com3} for $(r_1,r_2,r_3)$ and trajectories can be chaotic. Indeed, consider the harmonic oscillator and the state
\be
\phi^{3\mathrm{p}}_{\textrm{W}} = \phi^{\textrm{pol}}_{3,1}\phi^{\textrm{pol}}_{4,0}\phi^{\textrm{pol}}_{4,0}+\ee^{\ii\pi/3}\phi^{\textrm{pol}}_{4,0}\phi^{\textrm{pol}}_{3,1}\phi^{\textrm{pol}}_{4,0}+
\ee^{\ii\pi/7}\phi^{\textrm{pol}}_{4,0}\phi^{\textrm{pol}}_{4,0}\phi^{\textrm{pol}}_{3,1}.
\label{80}
\en
The Lyapunov exponent is approximately equal to $0.12$, which indicates chaos. A particular set of trajectories is shown in figure~\ref{2dharmonic_3part}.

\subsection{Superpositions with 3 single-particle basis states}\label{3bs}
In the previous section we have seen that the states \eqref{3part_GHZ} and \eqref{3part_W} do not lead to chaos for states using two states $\phi_i(x,y)$ of the form (\ref{5}). Chaos is possible for states that are formed using three such states. More precisely, we consider the states 
\begin{equation}
\phi_{3}^{\textrm{3p}}(\mathbf{x}_1,\mathbf{x}_2,\mathbf{x}_3)=c_1\phi_{1}(\mathbf{x}_1)\phi_{1}(\mathbf{x}_2)\phi_{1}(\mathbf{x}_3)+c_2\phi_{2}(\mathbf{x}_1)\phi_{2}(\mathbf{x}_2)\phi_{2}(\mathbf{x}_3)+c_3\phi_3(\mathbf{x}_1)\phi_3(\mathbf{x}_2)\phi_3(\mathbf{x}_3).
\label{3part_entangled}
\end{equation}

In the case of the harmonic oscillator, we consider
\begin{equation}
\phi_{3}^{\textrm{3p}} = \phi^{\textrm{2d}}_{3,1}\phi^{\textrm{2d}}_{3,1}\phi^{\textrm{2d}}_{3,1}+\ee^{\ii\pi/3}\phi^{\textrm{2d}}_{4,0}\phi^{\textrm{2d}}_{4,0}\phi^{\textrm{2d}}_{4,0}+\ee^{\ii\pi/7}\phi^{\textrm{2d}}_{2,2}\phi^{\textrm{2d}}_{2,2}\phi^{\textrm{2d}}_{2,2}.
\label{70}
\end{equation}
The Lyapunov exponent is approximately $0.19$, so that there is chaos.
\begin{figure}[ht]
\centering
\includegraphics{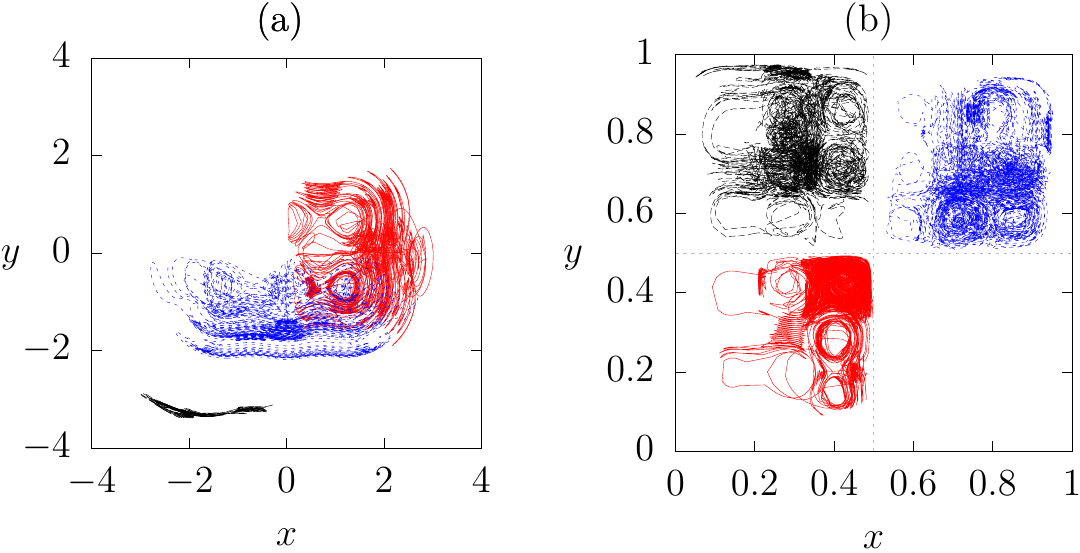}
\caption{ (a) and (b) Bohmian trajectories for a system of three particles respectively for the entangled wave functions (\ref{70}) and (\ref{71}), and with initial configuration respectively $(x_1,y_1,x_2,y_2,x_3,y_3)=(-2.98281, -1.92732, 2.84168, -0.12871, -2.43547, -0.292984)$ and $(x_1,y_1,x_2,y_2,x_3,y_3)=(0.383739, 0.882733, 0.464473, 0.481246, 0.616311, 0.586823)$. The black dashed curve corresponds to the trajectory of particle $1$, the red curve to that of particle $2$ and the blue one to that of particle $3$. 
}
\label{2dharmonic_3part_rect}
\end{figure}
Figure~\ref{2dharmonic_3part_rect}(a) shows a particular set of trajectories for this system.

As a second example, consider the square box and the state
\be
\phi_{3}^{\textrm{3p}} = \phi^{\textrm{box}}_{5,5}\phi^{\textrm{box}}_{5,5}\phi^{\textrm{box}}_{5,5}+ \ee^{\ii\pi/3}\phi^{\textrm{box}}_{7,1}\phi^{\textrm{box}}_{7,1}\phi^{\textrm{box}}_{7,1}
+\ee^{\ii\pi/7}\phi^{\textrm{box}}_{1,7}\phi^{\textrm{box}}_{1,7}\phi^{\textrm{box}}_{1,7}
\label{71}
\en
which has the same coefficients as the state (\ref{70}). A typical set of Bohmian trajectories is shown in figure~\ref{2dharmonic_3part_rect}(b). Just as in the two-particle case, the particles move in different quadrants of the box. The Lyapunov exponent is approximately $25$, so that there is chaos.

\subsection{Chaos and entanglement}\label{entanglementmeasure}
In this section we explore the relation between the average Lyapunov exponent and entanglement measures. Here, we consider three measures of multipartite entanglement: the Meyer-Wallach measure $Q$~\eqref{Q}, the geometric measure $E_G$\eqref{E_G} and the three-tangle $\tau_3$~\eqref{tau_3}. These entanglement measures vanish if the state $\psi$ is separable. In that case, the motion of the Bohmian particles is uncorrelated. Conversely, when the measure vanishes this implies that the state $\psi$ is separable in the case of the first two measures, but not in the case of the three-tangle.

These measures will not fully agree with the average Lyapunov exponent, as they do not depend on the choice of the basis states $|0\rangle$ and $|1\rangle$ (just like the participation ratio), while this influences the possibility of chaos, as we have seen before. In addition, note that entanglement is not sufficient to imply chaos. As a simple example, consider real wave functions. The wave function can be arbitrarily much entangled, but the Bohmian configuration is static, so that there is no chaos. 

In the following, we will consider stationary states belonging to the W entanglement class, i.e.\ states that are related by an invertible local operation to the W state $|\textrm{W}\rangle=(|001\rangle+|010\rangle+|100\rangle)/\sqrt{3}$. These states can be parametrized as follows
\begin{equation}
|\textrm{W}(a,b,c)\rangle=\sqrt{a}|001\rangle+\sqrt{b}|010\rangle+\sqrt{c}|100\rangle+\sqrt{1-(a+b+c)}|000\rangle,
\label{W-state}
\end{equation}
where $a,b,c\geq 0$, $a+b+c\leq 1$, and $|0\rangle$ and $|1\rangle$ are single-particle basis states~\cite{duer00}. We will take $\langle \mathbf{x}|0\rangle$ and $\langle \mathbf{x}|1\rangle$ to be the eigenstates $\phi^{\textrm{pol}}_{n_r,n_l}(r,\varphi)$ of the $2$-d harmonic oscillator. For these states, the three-tangle vanishes for all values of $a,b$ and $c$. Since the amount of chaos generically varies with the choice of coefficients, this measure will not correlate with the amount of chaos.

We consider first the following one-parameter family of states 
\be
|\textrm{W}(a,1/2-a,1/2)\rangle = \sqrt{a}|001\rangle+\sqrt{1/2-a}|010\rangle+\sqrt{1/2}|100\rangle,
\label{100}
\en
with $\langle \mathbf{x}|0\rangle=\phi^{\textrm{pol}}_{4,0}(\mathbf{x})$ and $\langle \mathbf{x}|1\rangle=\phi^{\textrm{pol}}_{3,1}(\mathbf{x})$ and $a\in[0,1/2]$, which have energy $15$. 
\begin{figure}[h]
\centering
\includegraphics{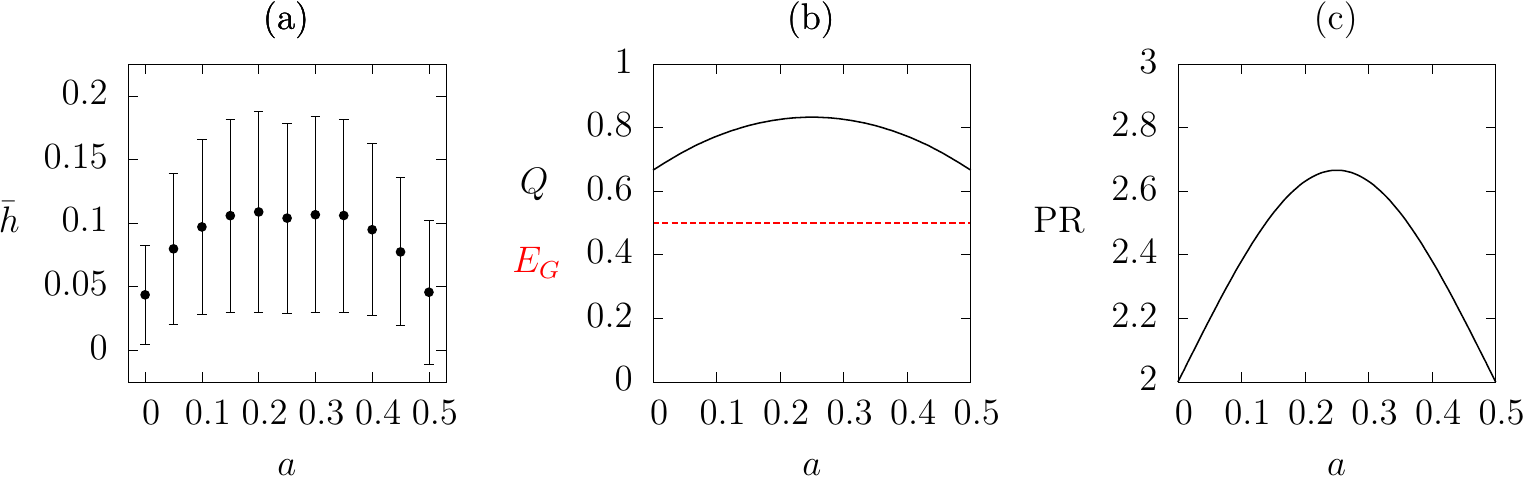}
\caption{Plots for the state \eqref{100}. (a) Average Lyapunov exponent $\bar{h}$ as a function of $a$. (b) Meyer-Wallach measure of entanglement ($Q$) (black) and geometric measure of entanglement ($E_G$) (red dashed) as a function of $a$. (c) Participation ratio for these states as a function of $a$.}
\label{b=0.5-a_c=0.5}
\end{figure}
Figure~\ref{b=0.5-a_c=0.5} shows from left to right the average Lyapunov exponent $\bar{h}$ (computed from 300 trajectories), the Meyer-Wallach and geometric measure of entanglement and the participation ratio as a function of the parameter $a$. For the states (\ref{100}), the three-tangle and the geometric measure of entanglement are independent of $a$; they have the value of $0$ and $1/2$ respectively. The average Lyapunov exponent $\bar{h}$ varies with $a$, is maximal at $a=1/4$ and minimal at $a=0$ and $a=1/2$. Both the participation ratio and the Meyer-Wallach measure of entanglement have the same general behavior as $\bar{h}$ with respect to $a$. This agreement in behavior can be understood by noticing that the states $|001\rangle$, $|010\rangle$ and $|100\rangle$ from which $|\textrm{W}(a,1/2-a,1/2)\rangle$ is constructed have similar spatial variations. For $a=0$ and $a=1/2$, $|\textrm{W}(a,1/2-a,1/2)\rangle$ has only two terms, with one particle decoupling from the others, which explains why the Lyapunov exponent is minimal for these values. 

\begin{figure}[htb]
\centering
\includegraphics{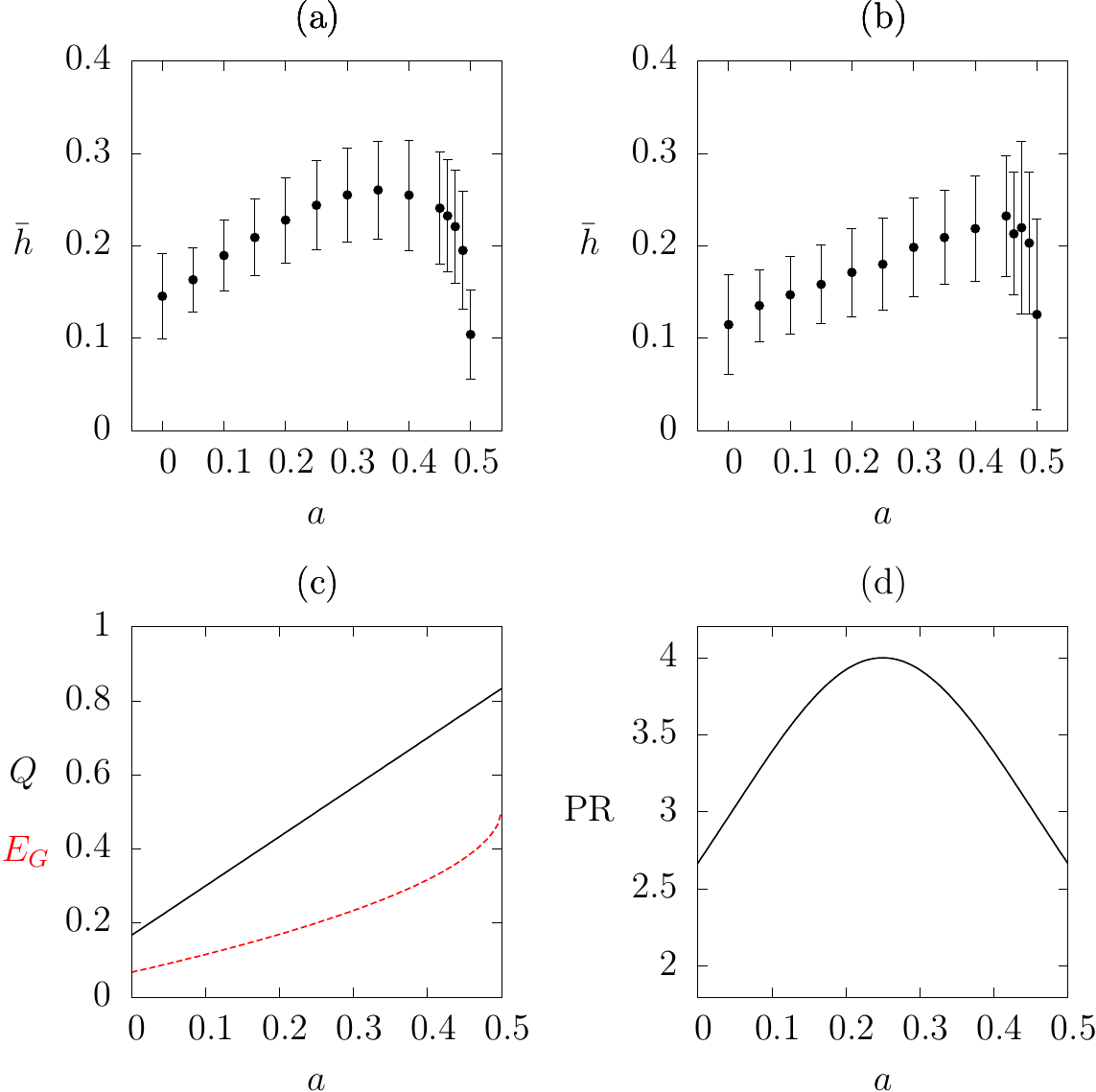}
\caption{(a) and (b) respectively display the Lyapunov exponent $\bar{h}$ for the states $|\textrm{W}(a,1/4,1/4)\rangle$ of energy $15$ and $|\textrm{W}'(a,1/4,1/4)\rangle$ of energy $21$ given by Eq.~(\ref{101}). (c) Meyer-Wallach measure of entanglement ($Q$) (black) and geometric measure of entanglement ($E_G$) (red dashed) as a function of $a$. (d) Participation ratio for these states as a function of $a$. $Q$, $E_G$ and PR are equal for both states. }
\label{b=0.25_c=0.25}
\end{figure}

Now consider the other one-parameter family of stationary states 
\be
|\textrm{W}(a,1/4,1/4)\rangle = \sqrt{a}|001\rangle+\sqrt{1/4}|010\rangle+\sqrt{1/4}|100\rangle+\sqrt{(1/2-a)}|000\rangle,
\label{101}
\en
with again $\langle \mathbf{x}|0\rangle=\phi^{\textrm{pol}}_{4,0}(\mathbf{x})$ and $\langle \mathbf{x}|1\rangle=\phi^{\textrm{pol}}_{3,1}(\mathbf{x})$ and $a\in[0,1/2]$, which have energy $15$. Figure~\ref{b=0.25_c=0.25}(a) shows $\bar{h}$, (c) the Meyer-Wallach and geometric measures of entanglement and (d) the participation ratio as a function of $a$. The lower value of $\bar{h}$ for $a=0$ may originate from the fact that the third particle decouples from the other ones. The lower value of $\bar{h}$ for $a=1/2$ may come from the fact that then the state is of the form \eqref{3part_W} and as mentioned before admits then one constant of motion. These properties do not exclude chaos but reduce the possible amount of chaos.
The agreement of $\bar{h}$ with the Meyer-Wallach and geometric measures of entanglement or the participation ratio is not as good as in the previous example. The Meyer-Wallach measure and the geometric measure of entanglement are monotonously increasing functions of $a$, unlike $\bar{h}$ which has a maximum. The participation ratio reaches a maximum, just like $\bar{h}$,  but its maximum is reached for a different value of $a$. The discrepancy between the Lyapunov exponent and the measures of complexity of the wave function originates from the fact that the former depends on the particular form of the terms in the superposition, while the latter do not. In this case, unlike in the previous example, not all terms have the same form due to the presence of the term proportional to $|000\rangle$.

As a last example, we consider the state \eqref{101} with $\langle \mathbf{x}|0\rangle=\phi^{\textrm{pol}}_{4,2}(\mathbf{x})$ and $\langle \mathbf{x}|1\rangle=\phi^{\textrm{pol}}_{3,3}(\mathbf{x})$, $a\in[0,1/2]$. We denote this state as $|\textrm{W}'(a,1/4,1/4)\rangle$. It has energy $21$, which is higher than the energy of $|\textrm{W}(a,1/4,1/4)\rangle$.
The Lyapunov exponent $\bar{h}$ for this state, plotted in figure~\ref{b=0.25_c=0.25}(b), qualitatively displays the same behavior as in the case of $|\textrm{W}(a,1/4,1/4)\rangle$. However, it is interesting to see that for most values of $a$, $|\textrm{W}(a,1/4,1/4)\rangle$ gives higher values of $\bar{h}$, despite the fact that this state has lower energy than $|\textrm{W}'(a,1/4,1/4)\rangle$. So, just as we have already seen before in section \ref{sup4modes}, increasing the energy of the system does not necessarily lead to a higher amount of chaos in the Bohmian trajectories. Similar fluctuations as a funtion of the mean energy where observed in \cite{wisniacki06}.

In summary, not all measures of entanglement lead to a reasonable agreement with the amount of chaos. For the family of states (\ref{W-state}), the three-tangle vanishes whereas the Bohmian trajectories display variable amount of chaos as a function of the parameters. Similarly, the geometric measure of entanglement of the states (\ref{100}) is constant while the Lyapounov exponent varies as a function of the parameter $a$. The best agreement is obtained for the Meyer-Wallach measure of entanglement. However, even this measure will not always be suitable. For example, for the state $\phi^{3\mathrm{p}}_{\textrm{GHZ}}$ (see Eq.~\eqref{3part_GHZ}), $Q(\phi^{3\mathrm{p}}_{\textrm{GHZ}})=2(1-|c_1|^4-|c_2|^4)$, while the Bohmian trajectories are regular for any $c_1,c_2$ as shown in section~\ref{2bs}. (Actually, also the three-tangle is non-zero in this case: $\tau_3(\phi^{3\mathrm{p}}_{\textrm{GHZ}})=4|c_1|^2|c_2|^2$.)

\section{Conclusion}
We considered stationary states for a single particle in a 3-d  harmonic potential and two and three entangled particles in a 2-d harmonic potential and 2-d box for which there is chaotic Bohmian motion. This shows that moving nodes are not necessary for chaos. Rather the overall complexity of the wave function is important. As such, this provides strong evidence that Bohmian trajectories are typically chaotic, whether the nodes are stationary or not. 

We also studied how well the amount of chaos, as measured by the average Lyapunov exponent, correlates with the participation ratio and different measures of entanglement (three-tangle, geometric measure of entanglement and Meyer-Wallach measure of entanglement). We found that these quantities often tend to correlate to the amount of chaos. However, they do not depend on the form of the basis states used to expand the wave function. In addition, the participation ratio does not depend on the relative phases of the expansion coefficients of the state in the chosen basis. Since the amount of chaos does depend on these factors, there cannot be full agreement between these measures and the amount of chaos.

\section{Acknowledgments}
WS carried out part of this work at the Institut de Physique Nucl\'eaire, Atomique et de Spectroscopie of the University of Liege, Belgium, with support from the Actions de Recherches Concert\'ees (ARC) of the Belgium Wallonia-Brussels Federation under contract No.\ 12-17/02. Currently WS acknowledges support from the Deutsche Forschungsgemeinschaft. Computational resources have been provided by the Consortium des \'Equipements de Calcul Intensif (C\'ECI), funded by the F.R.S.-FNRS under Grant No. 2.5020.11. We want to thank Detlef D\"urr and R\'emy Dubertrand for useful discussions and the referees for helpful comments.

\setcounter{section}{1}

\appendix
\section{Quantifying chaos}\label{chaos}
Many tools have been developed in order to probe and quantify chaos of a dynamical system~\cite{wimberger14}. Here, we explain the methods used in this paper for 3d Bohmian trajectories. Chaos means that the trajectories are highly sensitive to the initial position~\cite{lichtenberg92}. More precisely, there will be chaos if trajectories exponentially diverge within small times. That is, if for initial positions $\mathbf{x}_0$ and $\tilde{\mathbf{x}}_0$, initially separated by a small distance $d_0$, we have  
\begin{equation}
|\mathbf{d}(t)|\equiv|\mathbf{x}(t)-\tilde{\mathbf{x}}(t)|=d_0 \ee^{h(\mathbf{x}_0,\tilde{\mathbf{x}}_0)t},\;\;\;\;\;\;h(\mathbf{x}_0,\tilde{\mathbf{x}}_0)t \ll 1 ,
\end{equation}
with $h(\mathbf{x}_0,\tilde{\mathbf{x}}_0)>0$. 

The amount of chaos is quantified by the Lyapunov exponents~\cite{lichtenberg92}, given by
\begin{equation}
h(\mathbf{x}_0,\mathbf{e}_0)=\lim_{t\to\infty}\left({\lim_{d_0\to 0}{\frac{1}{t}\ln\left(\frac{|\mathbf{d}(t)|}{d_0}\right)}}\right),
\label{lyap_true}
\end{equation}
where $d_0$ is the initial distance between two trajectories starting at $\mathbf{x}(t_0)=\mathbf{x}_0$ and $\tilde{\mathbf{x}}(t_0)=\mathbf{x}_0+d_0 \mathbf{e}_0$ with $\mathbf{e}_0$ a unit vector and $|\mathbf{d}(t)|$ is the distance between these two positions as a function of time. While this quantity depends on $\mathbf{e}_0$, it can attain at most three different values (which corresponds to the number of dimensions of the configuration space). It can be shown that for almost all directions, i.e.\ almost all $\mathbf{e}_0$, the maximal value is obtained~\cite{lichtenberg92}. 

\begin{figure}[htb]
\centering
\includegraphics{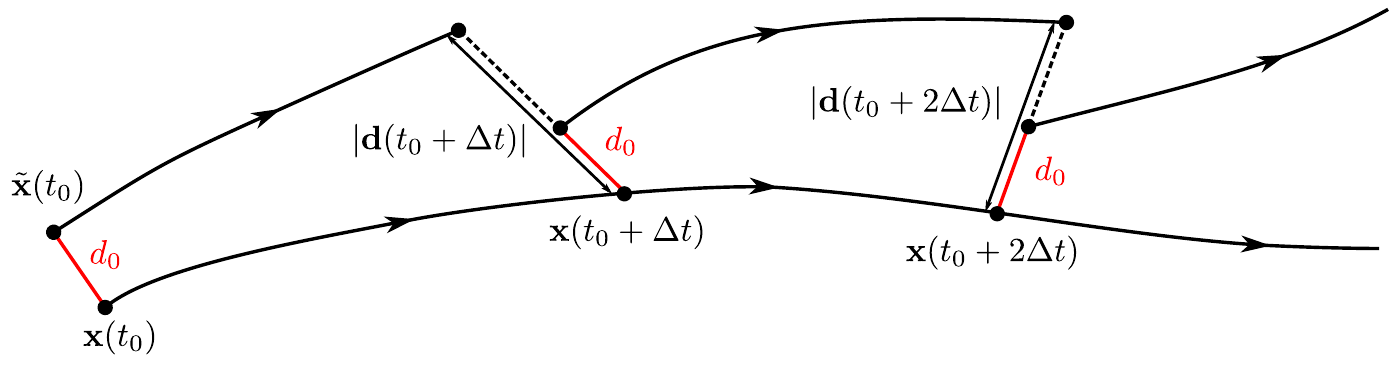}
\caption{In order to numerically compute the Lyapunov exponent, the distance between two initially close trajectories is rescaled every $\Delta t$.}
\label{lyap_comp}
\end{figure}

In order to numerically compute the Lyapunov exponent, we use the procedure~\cite{Benettin76,lichtenberg92} illustrated in figure~\ref{lyap_comp}. Starting from two initial positions $\mathbf{x}_0$ and $\tilde{\mathbf{x}}_0$ at time $t_0$ separated by a distance $d_0$, we numerically integrate the trajectories up to some time $t_0 + \Delta t$ to obtain ${\bf d}(t_0 + \Delta t) = \tilde{\mathbf{x}}(t_0 + \Delta t) - \mathbf{x}(t_0 + \Delta t)$. Then we repeat the procedure with initial positions $\mathbf{x}(t_0 + \Delta t)$ and $\mathbf{x}(t_0 + \Delta t) + d_0{\bf e}(t_0 + \Delta t)$, where ${\bf e}(t_0 + \Delta t)$ is the unit vector in the direction of ${\bf d}(t_0 + \Delta t)$. So the separation between these positions is rescaled to $d_0$ every $\Delta t$. We repeat this $N$ times.  
For large total time $T=N\Delta t$, the maximal Lyapunov exponent is then in good approximation given by 
\begin{equation}
h(\mathbf{x}_0,T)\approx \frac{1}{T}\sum_{k=1}^{N}{\ln\left(\frac{|\mathbf{d}(t_0+k \Delta t)|}{d_0}\right)}.
\end{equation}
A positive value of $h(\mathbf{x}_0)$ indicates that the dynamics is (strongly) chaotic, while $h(\mathbf{x}_0)\leq 0$ indicates that the trajectories stay close together in the sense that they do not diverge exponentially. 

The Lyapunov exponent defined in~(\ref{lyap_true}) quantifies the amount of chaos for particular trajectories. In order to probe the amount of chaos for an ensemble of trajectories all with the same wave function, one has to sample the initial position and, for every point, compute the Lyapunov exponent. Then one can either plot a map of these exponents or compute the spatial average. The former method may distinguish regions in space that lead to chaotic motion and regions that do not. However, for simplicity we just presented averages in this work.

While the discussion was so far restricted to a 3-dimensional system, it straightforwardly generalizes to a many-particle system whose configuration space is $n$-dimensional.

We benchmarked this method of computation of the Lyapunov exponent by computing $h(\mathbf{x}_0)$ for the H\'enon-Heiles Hamiltonian, as was also done in~\cite{Benettin76}. We also computed the Lyapunov exponent of Bohmian trajectories for the systems studied in references~\cite{wisniacki05,wisniacki06}. In both cases, our results were in good agreement. In order to assess the validity of our results, we checked the agreement between the Lyapunov exponents computed for different values of $d_0$, $\mathbf{e}_0$ and $\Delta t$.

There is another way to qualitatively probe the chaotic behavior of a dynamical system which consists in computing a Poincar\'e section, also called a section of surface~\cite{lichtenberg92}. The Poincar\'e section is obtained by considering the points of intersection of an $n$-d  trajectory $x(t)$ with a hyperplane of dimension $n-1$. Regions in the Poincar\'e section that display a regular pattern of points indicate a regular part of the trajectory. Regions where the points are densely and randomly distributed, indicate a chaotic part of the trajectory.

\section{Energy eigenstates for the considered systems}\label{energyeigenstates}
In this appendix, we recall the energy eigenstates and eigenvalues for the potentials we consider in the paper. In the 2-d  case these are the  harmonic oscillator potential and the square box potential. In the 3-d case this is the harmonic oscillator potential.

\subsection{2-d  systems}
\subsubsection{Square box potential}
For a particle moving in a square box with sides of length one~\cite{Cohen-Tannoudji91}, the energy eigenstates are  
\begin{equation}
\phi^{\textrm{box}}_{n_x,n_y}(x,y)=2\sin(n_x \pi x)\sin(n_y \pi y)
\label{eigenstate_box_2d}
\end{equation}
with the $n_i=1,2,\dots$ $(i=x,y)$ and have energy
\begin{equation}
E_{n_x,n_y}=\frac{\pi^2}{2 }\left(n_x^2+n_y^2\right).
\label{energy_box_2d}
\end{equation}
These eigenstates are real and there is some degeneracy. States $\phi_{n_1,n_2}$ and $\phi_{n_2,n_1}$ have the same energy. Moreover, there may be some accidental degeneracy as for example for the states $\phi^{\textrm{box}}_{5,5}$ and $\phi^{\textrm{box}}_{7,1}$, or $\phi^{\textrm{box}}_{7,4}$ and $\phi^{\textrm{box}}_{8,1}$.

\subsubsection{Harmonic oscillator potential}
For the harmonic oscillator~\cite{wisniacki05,Cohen-Tannoudji91}, the Hamiltonian is given by 
\begin{equation}
\widehat{H}_{\textrm{2d}}=\frac{1}{2}( \widehat{p}_x^2 + \widehat{p}_y^2 ) +  \frac{1}{2}  (\widehat{x}^2 + \widehat{y}^2) ,
\end{equation}
which simply corresponds to the sum of the Hamiltonians of two $1$-d  harmonic oscillators with the same frequency in the $x$ and $y$ direction. 

The eigenstates of $\widehat{H}_{\textrm{2d}}$ can be constructed as products of eigenstates of the $1$-d  harmonic oscillator and are given by
\begin{equation}
\phi^{\textrm{2d}}_{n_x,n_y}(x,y)=\frac{1}{\sqrt{\pi 2^{n_x+n_y}n_x!n_y!}}\ee^{-(x^2+y^2)/2}H_{n_x}( x)H_{n_y}( y),
\label{eigenstate_har_2dR}
\end{equation}
where $H_n(x)$ are the Hermite polynomials. These states have energy 
\begin{equation}
E^{\textrm{2d}}_{n_x,n_y}= n_x+n_y+1, 
\end{equation}
which is $(n_x+n_y+1)$-fold degenerate. The states $\phi^{\textrm{2d}}_{n_x,n_y}(x,y)$ are real.

Alternatively, one can use energy eigenstates that are also eigenstates of the $z$-component of angular momentum $\widehat{L}_z$~\cite{Cohen-Tannoudji91,Levin02}. In polar coordinates, the common eigenstates of $\widehat{H}_{\textrm{2d}}$ and $\widehat{L}_z$ are given by
\begin{equation}
\phi^{\textrm{pol}}_{n_r,n_l}(r,\varphi)=\mathcal{N}_{n_r,n_l} r^{|n_r-n_l|}\ee^{-r^2/2}L_n^{|n_r-n_l|}(r^2)\ee^{\ii(n_r-n_l)\varphi},
\label{eigenstate_har_2dC}
\end{equation}
with
\begin{equation}
\mathcal{N}_{n_r,n_l}=\sqrt{\frac{n!}{\pi(n+|n_r-n_l|)!}}
\end{equation}
and with $n= \textrm{min}(n_l,n_r)$. The energy eigenvalue is
\begin{equation}
E^{\textrm{pol}}_{n_r,n_l}= n_r+n_l+1
\end{equation}
and the $\widehat{L}_z$ eigenvalue is $(n_r-n_l)$ with $n_i=0,1,2,\dots$ $(i=r,l)$. Again, the energy $E^{\textrm{pol}}_{n_r,n_l}$ is $(n_r+n_l+1)$-fold degenerate.

\subsection{3-d  systems}

\subsubsection{Harmonic oscillator}
The Hamiltonian corresponding to a particle in a 3-d isotropic harmonic potential is 
\begin{equation}
\widehat{H}_{\textrm{3d}}=\frac{1}{2}(\widehat{p}_x^2 + \widehat{p}_y^2  + \widehat{p}_z^2)+\frac{1}{2}(\widehat{x}^2+\widehat{y}^2+\widehat{z}^2),
\end{equation}
which corresponds to the sum of the Hamiltonians of three $1$-d harmonic oscillators with the same frequency in the $x$, $y$ and $z$ direction. 

The eigenstates of $\widehat{H}_{\textrm{3d}}$ are simply product of eigenstate of the 1d harmonic oscillator~\cite{Cohen-Tannoudji91}
\begin{equation}
\phi^{\textrm{3d}}_{n_x,n_y,n_z}(x,y,z)=\frac{ \ee^{-(x^2+y^2+z^2)/2}H_{n_x}( x)H_{n_y}( y)H_{n_z}( z)}{\sqrt{\pi 2^{n_x+n_y+n_z}n_x!n_y ! n_z!}}
\label{eigenstate_har_3dR}
\end{equation}
where $H_n(x)$ are the Hermite polynomials. These states have energy 
\begin{equation}
E^{\textrm{3d}}_{n_x,n_y,n_z}=n_x+n_y+n_z+\frac{3}{2}.
\end{equation}
The energy $E^{\textrm{3d}}_{n_x,n_y,n_z}$ is $(n+1)(n+2)/2$-fold degenerate, with $n=n_x+n_y+n_z$. The states $\phi^{\textrm{3d}}_{n_x,n_y,n_z}(x,y,z)$ are real.

Alternatively, one can use states that are also eigenstates of $\widehat{\mathbf{L}}^2$ and $\widehat{L}_z$:
\begin{equation}
\phi^{\textrm{sph}}_{k,l,m}(r,\theta,\varphi)=\mathcal{N}_{k,l}\left(\frac{ r}{\sqrt{2}}\right)^{l} \ee^{- r^2/2}L_{k}^{l+1/2} \left( r^2 \right)Y_l^m(\theta,\varphi),
\label{eigenstate_har_3dsph}
\end{equation}
with
\begin{equation}
\mathcal{N}_{k,l}=\sqrt{\sqrt{\frac{1}{4\pi}}\frac{ 2^{2k+2kl+3}k!}{(2k+2l+1)!!}}.
\end{equation}
The energy is 
\begin{equation}
E^{\textrm{sph}}_{k,l}=n+\frac{3}{2},
\end{equation}
where $n=2k+l$. The $l$ quantum number can take the value $l=0,2,\dots,n-2,n$ if $n$ is even and $l=1,3,\dots,n-2,n$ if $n$ is odd, and $m=-l,-l+1,\dots, l-1,l$ \cite{robinett06}.

\bibliographystyle{unsrt}
\bibliography{ref.bib}

\end{document}